\documentclass[journal]{IEEEtran}
\usepackage[algo2e]{algorithm2e}
\usepackage{amssymb}
\usepackage{amsmath, amsthm, amsfonts}
\usepackage{bbm}
\usepackage{algorithm}
\usepackage{ wrapfig, setspace, booktabs}
 \usepackage[pdftex]{graphicx}
\usepackage{subcaption}
\usepackage[T1]{fontenc}
\usepackage{fancyhdr}
\usepackage{lastpage}
\usepackage{url, lipsum}
\usepackage{mdwmath}
\usepackage{adjustbox}
\usepackage{authblk}
\usepackage{soul}
\usepackage{pifont}
\usepackage{mathtools}
\usepackage{textcomp}
\usepackage{mathabx}
\usepackage[mathscr]{euscript}
\DeclareMathOperator{\sinc}{sinc}
\DeclareMathOperator{\sigmoid}{sigmoid}
\usepackage{color}
\usepackage[dvipsnames]{xcolor}
\ifCLASSINFOpdf
\else
\fi

\newcommand{\mycomment}[1]{}
\begin{document}

\title{Time-Frequency Warped Waveforms \\ for Well-Contained Massive Machine Type Communications}

\author{Mostafa Ibrahim*,~\IEEEmembership{Student~Member,~IEEE}, Huseyin Arslan,~\IEEEmembership{Fellow,~IEEE}, Hakan Ali Cirpan ,~\IEEEmembership{Member,~IEEE}, and Sabit~Ekin,~\IEEEmembership{Senior Member,~IEEE}

 % <-this % stops a space
 \thanks{Copyright (c) 2023 IEEE. Personal use of this material is permitted. However, permission to use this material for any other purposes must be obtained from the IEEE by sending a request to pubs-permissions@ieee.org.This paper has been accepted by IEEE JSAC special issue on 3GPP Technologies: 5G-Advanced and Beyond. Copyright may be transferred without notice, after which this version may no longer be accessible.
 (\textit{*Corresponding author: Mostafa Ibrahim}.)}
\thanks{Mostafa Ibrahim,  is with the School of Electrical and Computer Engineering, Oklahoma State University, Oklahoma, USA (E-mail: mostafa.ibrahim@okstate.edu).}
\thanks{Huseyin Arslan  is with the Department of Electrical and Electronics Engineering, Istanbul Medipol University, Istanbul, Turkey  (E-mail: huseyinarslan@medipol.edu.tr
).}
\thanks{Hakan Ali Cirpan   is with the Department of Electronics and Communication Engineering, Istanbul Technical University, Istanbul, Turkey (E-mail: hakan.cirpan@itu.edu.tr)}
\thanks{Sabit Ekin  is with the Departments of Engineering Technology, and Electrical \& Computer Engineering, Texas A\&M University, Texas, USA (E-mail: sabitekin@tamu.edu).}}

\maketitle

\begin{abstract}
This paper proposes a novel time-frequency warped waveform for short symbols, massive machine-type communication (mMTC), and internet of things (IoT) applications. The waveform is composed of asymmetric raised cosine (RC) pulses to increase the signal containment in time and frequency domains. The waveform has low power tails in the time domain, hence better performance in the presence of delay spread and time offsets. The time-axis warping unitary transform is applied to control the waveform occupancy in time-frequency space and to compensate for the usage of high roll-off factor pulses at the symbol edges. The paper explains a step-by-step analysis for determining the roll-off factors profile and the warping functions. 
Gains are presented over the conventional Zero-tail Discrete Fourier Transform-spread-Orthogonal Frequency Division Multiplexing (ZT-DFT-s-OFDM), and Cyclic prefix (CP) DFT-s-OFDM schemes in the simulations section.
\end{abstract}

% Note that keywords are not normally used for peer review papers.
\begin{IEEEkeywords}
6G and Beyond, Waveform, Time-Frequency Warping, massive machine-type communication, mMTC, IoT.
\end{IEEEkeywords}

\IEEEpeerreviewmaketitle

% is described when applied to the time-frequency plane
\section{Introduction}
\IEEEPARstart{T}{he} number of internet of things (IoT) connected devices is expected to grow exponentially in the next decade serving various use cases with highly diverse requirements. Machine Type Communication (MTC), with its massive and low latency unlimited wireless connectivity, is a key driver in this expanding trend. To support this massive MTC (mMTC) vision, it is required to design ultra-low power receivers, highly efficient sleep modes, and ultra-low-cost MTC devices. This paper uses the concept of time-frequency warping \cite{Ibrahim} as a waveform modifier on signals intended for relaxed synchronization, energy efficiency, and low latency scenarios \cite{schulz2017latency}. Some of the mMTC use cases are swarm networking transport systems, future factories, smart cities, zero-energy IoT, and smart contracts of distributed ledgers \cite{mahmood2021machine}. In such applications, synchronicity and orthogonality constitute a challenge due to sporadic access in fast dormancy operations \cite{6736749}. The key requirements for sixth-generation wireless (6G) and beyond massive machine and IoT waveforms are time localization, spectral confinement, very short packet transmission, low power consumption, and relaxed synchronization \cite{Zaidi,8367785}.  We propose a waveform that satisfies these requirements and gains performance over conventional waveform candidates.   

This paper is an extension of \cite{Ibrahim} that proposes a method to create a spectrally efficient well-contained symbol in time and frequency domains. The idea is to use high roll-off factors for the raised-cosine (RC) shaped pulses near the edges of the symbol and low roll-off factors for the inner pulses. This is because, in a zero-tail DFT-s-OFDM-like symbol, the outer pulses contribute more to the time domain zero tails than the internal pulses. This high containment enhances the performance in the presence of time dispersion or time and frequency offsets because the power leaked from the edge symbols would be minimum in the time domains. Then, the pulses occupancy in the frequency domain is controlled by introducing time-axis warping, which will align the spectral occupancy of all the pulses in the frequency domain. As a result, intersymbol interference (ISI) in the time and frequency domains is reduced.

Our preliminary work in \cite{Ibrahim} did not show how to choose the roll-off factors profile nor gave any guidance on the warping function determination. In this study, we attempt to introduce what is missing in that study but for SC-OFDM-based waveforms instead of OFDM-based waveforms. We propose methods to determine an optimized rolls-off factor $\alpha$ profile, followed by warping function determination that fits the decided $\alpha$ profile. The paper also explains some concepts related to modifying the warping function to make a more confined waveform.   

The time-frequency warping modifier is used in the fields of signal analysis and wavelet theory. In \cite{668543}, Evangelista extends the definition of dyadic wavelets to include frequency warped wavelets. Moreover, the time-frequency plane's flexible orthogonal and non-orthogonal tilings were defined using frequency warped wavelets. In \cite{8631608}, designing variable warped filters is proposed to fit the frequency response of a signal. Also, for electrocardiogram (ECG) pattern recognition, several works used warped time-frequency space to analyze the ECG signals \cite{1571142,6891378}. Time warping analysis is also used successfully in speech recognition \cite{467271}, room acoustics \cite{5946402}, and ultrasonic guided-wave characterization \cite{4803296}.

 The base symbol that we will use is Zero Tail DFT spread OFDM (ZT-DFT-s-OFDM) \cite{Berardinelli}; a variation of the (DFT-s-OFDM) modulation, aiming to decrease the zero tails power. DFT-s-OFDM is used as the modulation scheme for Long-Term Evolution (LTE) uplink schemes because of its low peak-to-average power ratio (PAPR), making it suitable for low-power systems. On top of that, it will be ideal for the proposed time-frequency warping method for ease of modulation. There are different variations of DFT-s-OFDM \cite{demir2019waveform}; a Unique Word OFDM (UW-OFDM) replaces the cyclic prefix (CP) with unique symbols in the time domain. ZT-DFT-s-OFM, on the other hand, replaces CP with zero symbols.

%mention the roll-off factors
In this paper, the pulses used for modulation can be symmetric or asymmetric RC pulses, with different roll-off factors on each side. As will be shown, asymmetric RC pulses allow for using low-power tails only where they are needed (at the edges). The asymmetric windowing and pulses were used in  \cite{7848842,6980135} to reduce out-of-band emissions and increase spectral efficiency.

Our contributions and key findings of this study are as follows:
\begin{itemize}
    \item Extending time-frequency warping concept to SC-OFDM schemes to fit future requirements of highly contained, energy-efficient waveforms. Zero tail DFT-s-OFDM is used as the basic waveform shape in this study because of its sinc-shaped pulses that we replace with the generalized raised-cosine shaped pulses with different roll-off factors ($\alpha$). The usage of RC pulses ensures orthogonality between pulses with various roll-off factors.
    \item A roll-off factor profile analysis is performed using the diminishing marginal utility approach, and different utility functions are explored. The analysis aims at deriving an optimized roll-off factors profile. 
    \item Insights on different concepts regarding the time-frequency occupancy of a warped pulse and its relationship with the warping function are given. We show that the first derivative of the warping function and the positions of pulses relative to the warping function affect the time-frequency occupancy of the pulse and, consequently, the whole symbol.
    \item For further spectral containment, we propose an asymmetrically shaped time-frequency warped raised-cosine pulse. The pulse has different roll-off factors ($\alpha$ s) on its sides; a higher $\alpha$ value is used towards the edge of the symbol, and a lower $\alpha$ value is used towards the center of the symbol.
    \item An analysis for determining the warping function based on a pre-planned roll-off factors profile is presented. The analysis uses linear programming to solve a multi-dimensional problem that cannot be solved analytically. 
    \item Transmitter and receiver schemes are proposed with implementation suggestions for computational complexity reduction. A discussion of system-level optimization using reinforcement learning (RL) and parametric trends is introduced. 
\end{itemize}

The rest of this paper is organized as follows; the time-frequency warping theory is discussed in Section II. Roll-off factors profile determination is discussed in Section III, followed by the warping function concepts and design in Section IV. The transceiver, modulator, and demodulator scheme designs of the proposed waveforms are discussed in Section V. Finally, an evaluation of the time-frequency waveform and comparison with the conventional counterpart waveforms is provided in Section VI.

\section{Warped Waveform Formulation}
In this section, axis warping unitary theory is briefly presented, and its effect is described when applied to the time-frequency plane.

\subsection{Axis Warping Theory}
 
 % more info can be added here
Axis warping transformation is a subclass of unitary transformations. The unitary operator $\mathbf{U}$ is a linear transformation that maps between two Hilbert spaces (i.e., $\mathbf{U} ~: ~\mathbf{L}^2(\mathbb{R}) \longmapsto  \mathbf{L}^2(\mathbb{R})$). Unitary transformations have the following characteristics. They maintain inner products (i.e., $\langle \mathbf{U} s, \mathbf{U} h\rangle = \langle s,h \rangle$), and they preserve energy (i.e., $ \lVert \mathbf{U} s \rVert ^2 = \lVert  s \rVert ^2 $). Axis warping is applied through the warping function $w$ that maps the axis $x$ to the new axis $w (x)$: 
\begin{equation}
w : \mathbb{R} \mapsto \mathbb{R}, ~~ x \mapsto w(x)  ~,
\end{equation}
where $w$ is a one-to-one, monotonic function. The warped orthogonal axes  $\Tilde x$, and $ \Tilde y$ are related to the original axes $ x$ and $ y$ through:
\begin{equation}
    \Tilde{x}=w(x), ~~ \Tilde{y}=y~ \dot m (w(x))~,
\end{equation}
where $m = w^{-1}$ is the inverse function of $w$. 
Axis warping of waveform $s$ expressed as follows  \cite{Baraniuk}:

\begin{equation}
    [\mathbf{U} s](x)= \lvert \dot w (x)  \rvert^{1/2} s[w(x)]~,
\end{equation}

where $\dot w$ represents the first derivative of the function $w$. The warping transform allows for non-uniform axes manipulation and provides a flexible method for controlling time-frequency occupancy.

\begin{figure}[ht]
\centering
    \includegraphics[width=.49\textwidth]{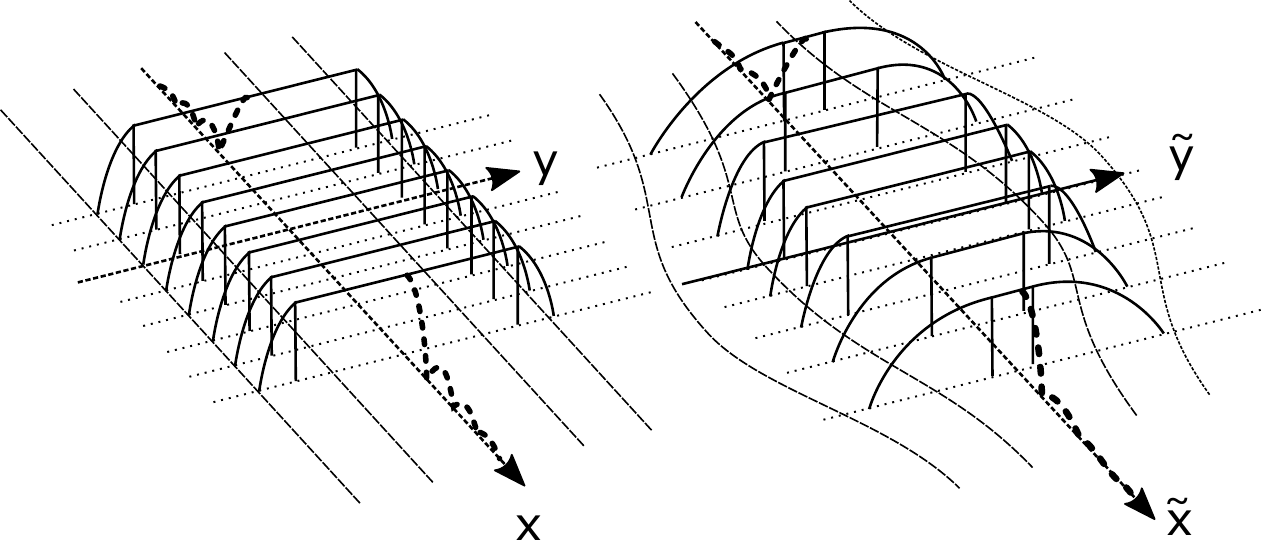}
    %\begin{center}
    \caption{Windowing versus Axis Warping with gradually changing roll-off factors.}
    \label{fig:concpt}
\end{figure}

In the previous study \cite{Ibrahim}, frequency axis warping was used to dilate the higher roll-off factors subcarriers at the edges of the OFDM-based symbol band,  to decrease the out-of-band emissions (OOBE) without losing spectral efficiency. This was the method to ensure that all the subcarriers had the same time domain occupancy.
In this paper, the context is massive MTC-type communication with low PAPR and high containment in the time-frequency plane. Therefore, we propose applying the time axis warping for an SC-OFDM-based waveform. Here we use axis warping to dilate the time domain pulses at the symbol edges to equalize the effect of the pulses having higher roll-off factors at the symbol edges. The warping concept is shown in Fig.\ref{fig:concpt} and its comparison with filtering in a generic x-y plane, where x and y can be used to represent time or frequency interchangeably. 

In the case of waveform filtering, the shape of the pulse is similar for all pulses. The spacing between the pulses on the x-axis is uniform, and the occupancy in the frequency domain is similar for all of them. For the warped waveform case, the pulse shape on the symbol's edges has higher roll-off factors, hence, lower tails on the x-axis. The pulses with higher roll-off factors have higher occupancy on the y-axis. Therefore, warping is used to shrink the x-y plane on the y-dimension by dilating the x-dimension at the position of the high $\alpha$ pulse. The warping function has to be designed to fit the roll-off factors of the pulses, and the roll-off factors should be chosen to have low power of tails without losing spectral efficiency. 

% In \cite{Ibrahim}, warping was applied to Orthogonal Frequency Division Multiplexing (OFDM) based symbols to decrease the out-of-band emissions (OOBE) without losing spectral efficiency. It was proposed to use raised cosine (RC) shaped sub-carriers with high roll-off factors at the edge sub-bands and to apply low roll-off factors for the inner sub-bands. Then, the warping modifier was used to unify the time domain occupancy for all sub-carriers.
%Moreover, the shapes of the pulses are different from the symmetric sub-carriers in \cite{Ibrahim}.
\subsection{Frequency Domain Deformation}
Axis warping in the time domain creates deformation in the frequency domain \cite{caporale2009design}. Let $H(f)$ be the frequency domain response of the unwarped version of the studied waveform. The frequency domain warping deformation results from the operator: 
\begin{equation}
\textbf{W}_\textbf{t} = \textbf{F}  \textbf{W}  \textbf{F}^\dagger , 
\end{equation}  
where $\textbf{F}$, and $\textbf{F}^\dagger$ are the Fourier and inverse Fourier transforms, respectively. $\textbf{W}$ is warping operator such that; $[\textbf{W}h](t)= \sqrt{\dot w (t)}    h[w(t)]$.
The first part of the $\textbf{W}_\textbf{t}$ operator is
\begin{equation}
\begin{aligned}
\textbf{W}   \textbf{F}^\dagger (t,f) & = \int_{\mathbb{R}} \sqrt{\dot w (t)}   \delta(w(t) - \tau)  e^{j 2 \pi f \tau} d\tau \\
& = \sqrt{\dot w (t)}   e^{j 2 \pi f w(t)} .
\label{eq:sqrt}
\end{aligned}
\end{equation}

Then,
\begin{equation}
\textbf{F}  \textbf{W}  \textbf{F}^\dagger (f,\nu)  = \int_{\mathbb{R}} \sqrt{\dot w (t)}   e^{j 2 \pi \nu w(t)}   e^{-j 2 \pi f t} dt .
\end{equation} 

So, the new shape of $H$ after warping is found by
 \begin{equation}
[\textbf{F}  \textbf{W}  \textbf{F}^\dagger H] (f) =   \int_{\mathbb{R}} H(\nu)  \int_{\mathbb{R}} \sqrt{\dot w (t)}   e^{-j 2 \pi( ft-\nu w(t))}  dt   d\nu  .
\label{eq:eq7}
 \end{equation}

The output of this operation can not be found analytically for warped RC pulses, or windows \cite{cook2012radar}. Therefore in the following sections, numerical methods will be used to calculate the frequency domain leakage based on the previous formulation.

\section{Roll-Off Factors Profile Determination }
In this section, we build our analysis based on RC pulses.
% In the following paragraphs, frequency out-of-band leakage, is exchangeable with time-domain tail power leakage. 
%%Side-lobes of the RC pulses at the edge of the band contribute most of the out-off-band power, and as far as we go away from the edge, this contribution gets less. Therefore, it is intuitive to have pulses with high $\alpha$ near the edges, as a high roll-off factor translates into lower side-lobes. On the other hand, the pulses away from the edge do not need to have high $\alpha$ values. %Let us assume RC pulses in the x domain at the edges of a band, as shown in Fig. \ref{fig:concpt}. All the side-lobes of the pulses add up, causing the out-of-band power. 
The side-lobes of the RC pulses at the band's edge contribute the majority of the out-of-band power, and as we move away from the edge, this contribution decreases. As a result, having pulses with higher $\alpha$ near the edges makes sense, because a high roll-off factor translates into lower side lobes. The pulses away from the edge, on the other hand, do not need to have high $\alpha$ values.

%%When the first pulse on the edge has a roll-off factor $\alpha_1$, the side-lobes of this pulse have the highest suppression. To keep the pulse orthogonal with other pulses in the band and to occupy the same window in the transform domain y, the wrapping function should extend the x-axis with a factor of  $(1+\alpha_1)$. This extension is a cost that is paid to get side-lobes suppression. 
When the first pulse on the edge has a roll-off factor of $\alpha_1= 1$, the side-lobes of this pulse are suppressed the most. The warping function should extend the x-axis by a factor of $(1+\alpha_1)$ to occupy the same window in the transform domain y. This extension is a price to pay for side-lobe suppression.
%%All the pulses near the edge contribute to the out-of-band power with different weights; therefore, they should all get suppressed side lobes. Consequently, an extension cost will be paid for each of them. This extension adds up at the end, causing a loss in spectral efficiency. Hence this expansion must be treated carefully to get the most out-of-band suppression with the least warping expansion.
Because all of the pulses near the edge contribute to out-of-band power (or zeros time gap) with different weights, they should all have suppressed side lobes with different $\alpha$ values. As a result, each of them will incur an extension cost. This extension adds up, resulting in a loss of spectral efficiency. As a result, this expansion must be handled with care in order to achieve the greatest out-of-band suppression with the least amount of warping expansion.
%%By moving further from the edge, pulses start to contribute less to the out-of-band power; hence it would be a loss to assign high $\alpha$ values for those pulses. This is precisely a case of Diminishing Marginal Utility. As an economics concept, diminishing marginal utility states that, as we spend more of a factor, which is, in our case, the roll-off factor of the pulses, the marginal utility diminishes. In our case, the out-of-band power reduced contribution for each pulse is what diminishes.
It would be inefficient to assign high $\alpha$ values for those pulses as they move further away from the edge because they start to contribute less to the out-of-band power. This is precisely a case of Diminishing Marginal Utility \cite{gossen1983laws}. According to the economic principle of diminishing marginal utility, as we spend more of a cost—in this case, the roll-off factor of the pulses—the marginal utility decreases. What decreases in our case is the reduced out-of-band power contribution for each pulse.
%%Following this analogy, roll-off factors will be chosen according to the law of equalizing marginal utility. We are choosing the utility as the suppression of the first out-of-band side-lobe and the cost of the expansion resulting from the warping function. 
Roll-off factors will be chosen in accordance with the law of equalizing marginal utility. We define the utility as the suppression of the first out-of-band side lobe and the cost as the warping function expansion.
%%The law of equalizing marginal utility in our case will translate into; the expansion that is paid for suppressing the $n^{th}$ lobe of the $n^{th}$ pulse should be the same as the expansion paid for suppressing the first lobe of the first pulse. It goes on like this for the rest of the pulses.
The expansion paid for suppressing the $n^{th}$ lobe of the $n^{th}$ pulse should be the same as that paid for suppressing the first lobe of the first pulse. 
\begin{equation}
    \dfrac{u_1}{c_1}=\dfrac{u_2}{c_2}= ... = \dfrac{u_n}{c_n}.
\label{eq:3}
\end{equation}
%%There can be different criteria for representing $u_n$ and $c_n$. In this study, we will propose three different criteria and solve them. $u_n$ represents the power suppression of the first lobe after the last pulse (The edge). 

Different criteria can be used to represent $u_n$ and $c_n$. In this study, we will propose and solve three different criteria. $u_ n$ denotes the power suppression of the first lobe just after the last pulse (The edge of the symbol).
%%The first lobe has the highest power of the side lobes. Assuming the worst case when all the pulses constructively interfere, the total power under the first lobe $P_T$ can be represented as. 
The power of the first lobe is the greatest of the side lobes. The worst-case scenario is when all the side lobes of the pulses constructively interfere. Assuming the worst case, the total power under the first lobe $P_T$ is represented as.
\begin{equation}
    P_T(x)= ( |L_1 (x) | + |L_2 (x) |+ ... +|L_n (x) |)^2 ,
\end{equation}
where $L_n(x)$ represents the $n^{th}$ lobe amplitude value of the $n^{th}$ pulse at $x$. Then, the total power under the first lobe is:
\begin{equation}
    P_T(x)=  \int_{1^{st} lobe} ( |L_1 (x) | + |L_2 (x) |+ ... +|L_n (x) |)^2  dx .
\end{equation}

%%The power suppression due to shaping the n^{th} side-lobe is the subtraction of the total side-lobes power before and after shaping the $n^{th}$ one.
The power suppression caused by shaping the $n^{th}$ side-lobe is calculated by subtracting the total power of the side-lobes before and after shaping the $n^{th}$ one on the duration of the first lobe after the edge.
\begin{equation}
  \begin{aligned}
   u_{n}=  \int_{1^{st} lobe} & ( |L_1 (x) | + |L_2 (x) |+ ... +|L_n (x) |)^2- \\
   & ( |L_1 (x) | + |L_2 (x) |+ ... +|L_{sn} (x) |)^2  dx .
     \end{aligned}
\end{equation}

 This can be reduced to:
\begin{equation}
  \begin{aligned}
   u_n=  \int_{1^{st} lobe} &(|L_{n} (x)|^2-|L_{sn} (x)|^2)+ \\ 
   & 2 ( |L_{n} (x)|-|L_{sn} (x)|)\sum_{m \neq n} |L_m(x)|  dx .
     \end{aligned}
     \label{eq:7}
\end{equation}

We will consider three different assumptions to solve Eq. (\ref{eq:3}) and Eq. (\ref{eq:7}) for RC pulses.  
\subsubsection{Case 1}
The the cost is represented as the -to be applied- equivalent warping expansion $c_n = \dfrac{1}{1+\alpha_n}$. $u_n$ is taken from Eq. (\ref{eq:7}). The equations are solved iteratively, and the solution of the roll-off factor profile starting with $\alpha_1=1$ is shown in Fig. \ref{fig:rollprofile} for 6 and 12 pulses.

\subsubsection{Case 2}
Here we define a simpler utility function as the suppression due to shaping the $n^{th}$ side-lobe by subtracting only the $n^{th}$ pulse side-lobe power before and after shaping.
\begin{equation}
 u_n=  \int_{1^{st} lobe} (|L_{n} (x)|^2-|L_{sn} (x)|^2 dx .
 \label{fig:smplutlty}
  \end{equation}
  
 The warping expansion is the same: $c_n = \dfrac{1}{1+\alpha_n}$, and $\alpha_1=1$. Results of solution 2 are shown in Fig. \ref{fig:rollprofile} for 6 and 12 pulses.
 
 \subsubsection{Case 3}
The assumption, in this case, is to contribute all the pulses similar to the symbol's $1^{st}$ lobe power. 
\begin{equation}
  \int_{1^{st} lobe} |L_{1} (x)|^2 dx =  ...  \int_{1^{st} lobe} |L_{n} (x)|^2 dx  .   
 \label{eq:case3}
  \end{equation}
  
\begin{figure}[h]
\centering
    \includegraphics[width=.48\textwidth]{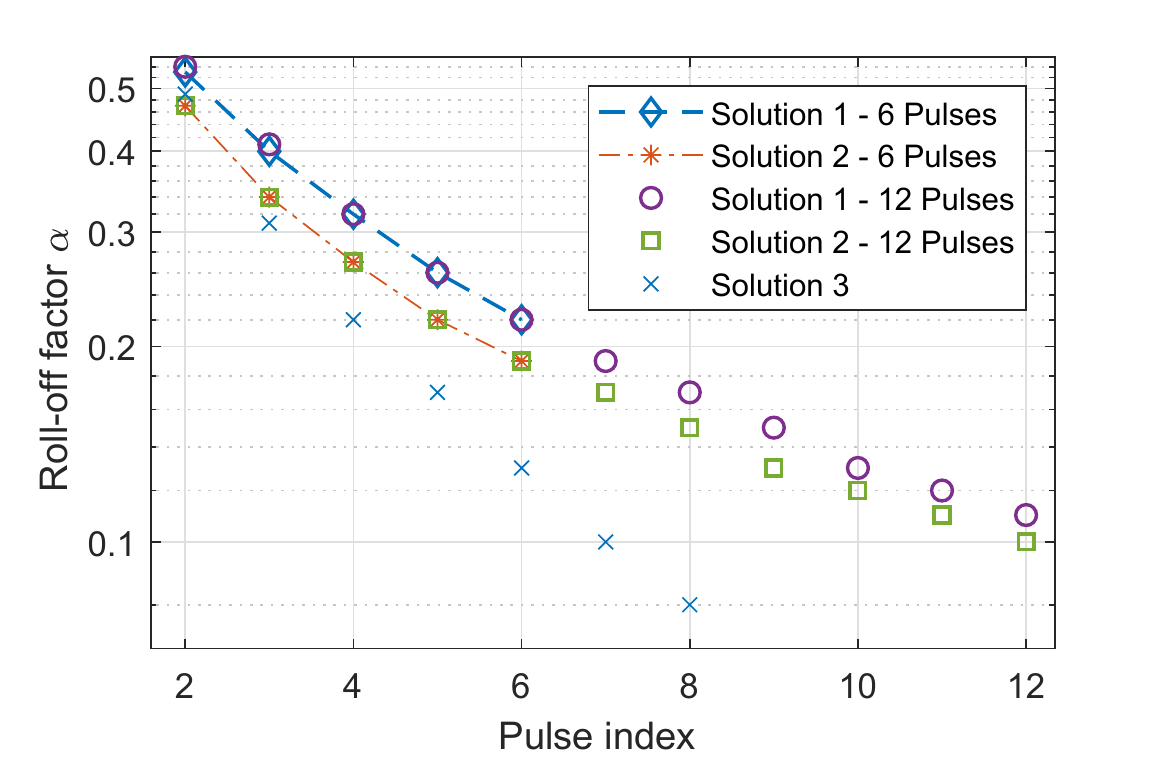}
    %\begin{center}
    \caption{Roll-off factors profile with pulse 1 has $\alpha = 1$.}
    \label{fig:rollprofile}
\end{figure}

This will give a steeper roll-off profile, as shown in Fig. \ref{fig:rollprofile}.
We can observe that different $u_n$ and $c_n$ criteria can result in different trends in the $\alpha$ profiles. Each will lead to a different zero-tails suppression, and different spectral efficiency, as will be shown in the next sections, The optimality of the roll-off factor profile will depend on the containment versus spectral efficiency compromise.  
Next, we discuss the concepts of designing the warping function based on a known roll-off factors profile. 
\section{Warping Function Design}
This section discusses the concepts for designing the warping function with the pulse shapes to achieve a well-contained waveform. The first idea is the slope of the warping profile; the second is about the pulse position with respect to the warping function. The third concept is the proposition of asymmetrical pulses.
\subsection{Introductory Concepts}

\subsubsection{Concept 1: The effect of warping function slope profile}

\begin{figure}[ht]
\centering

\begin{subfigure}{0.35\textwidth}
\includegraphics[width=1\linewidth]{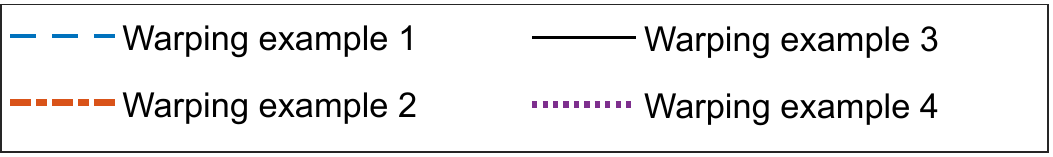} 
\label{fig:Alegend}
\end{subfigure}

    \begin{subfigure}{0.45\textwidth}
    \includegraphics[width=1\linewidth]{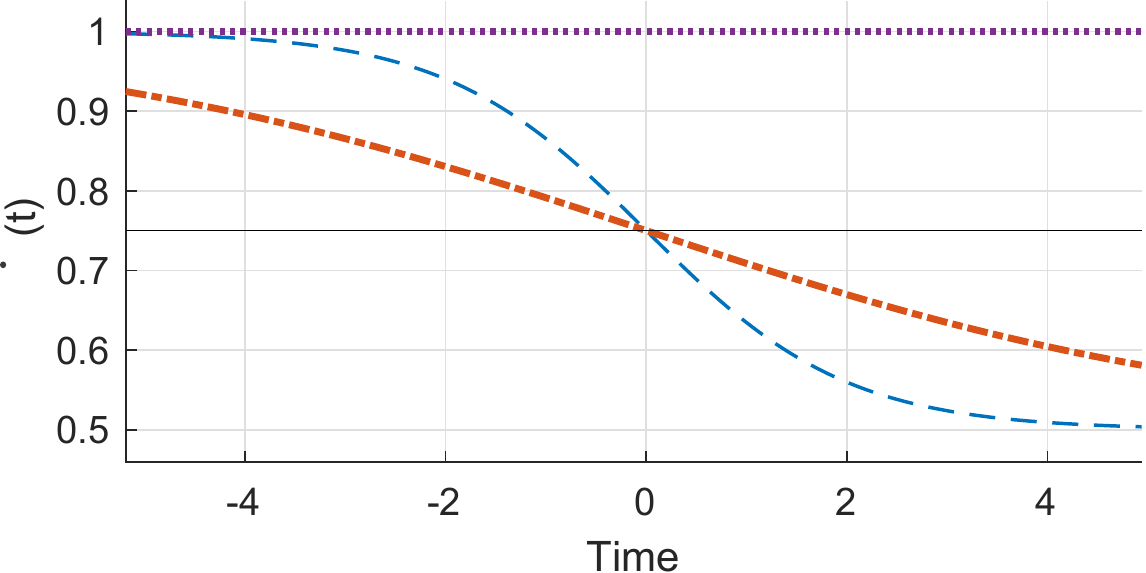} 
    \caption{Warping function slope.}
    \label{fig:figAw}
    \end{subfigure}
\begin{subfigure}{0.45\textwidth}
\includegraphics[width=1\linewidth]{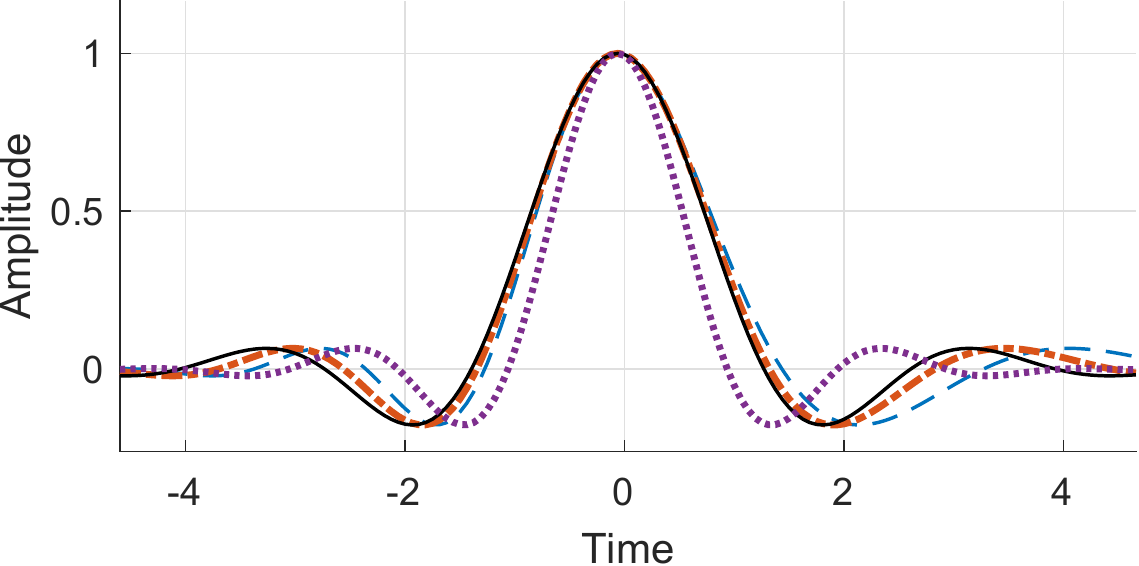} 
\caption{Time domain pulse.}
\label{fig:figAt}
\end{subfigure}
    \begin{subfigure}{0.45\textwidth}
    \includegraphics[width=1\linewidth]{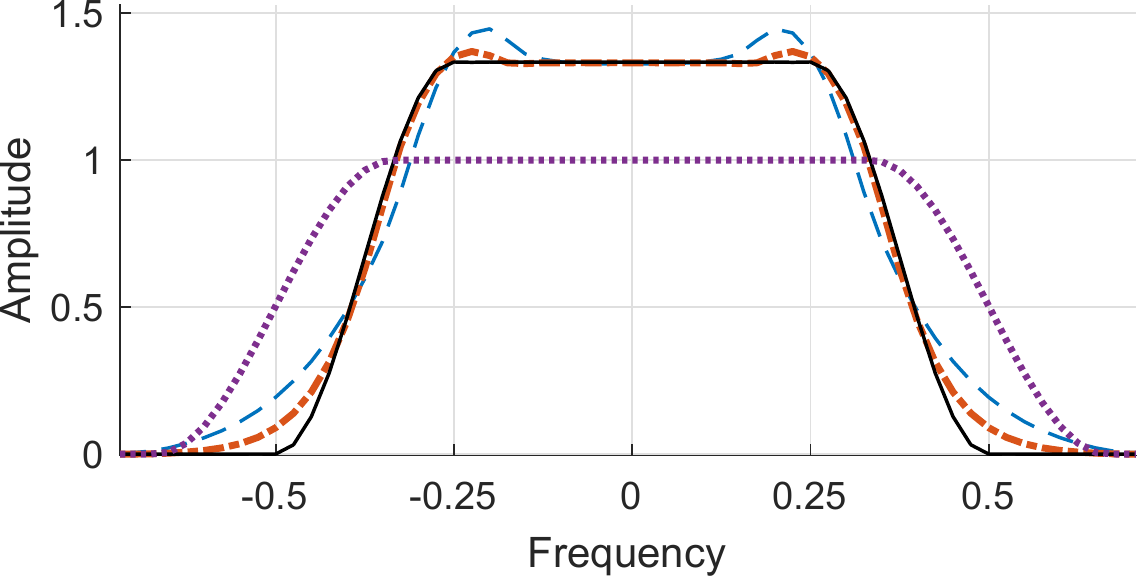} 
    \caption{Frequency domain window.}
    \label{fig:figAf}
    \end{subfigure}

\caption{Warping concept 1: different warping slopes, same roll-off factor 0.33.}
\label{fig:figA}
\end{figure}

The first derivative of the warping function at point x is proportional to the inverse of the expansion in the x-axis direction. In the time-domain pulses, the expansion is in the time domain. In Fig. \ref{fig:figA}, there are four examples of the warping functions first derivatives $\dot{\omega} (t)$. The RC pulse has a roll-off factor of $\alpha=0.33$. By replacing $t$ with $\omega(t)$, the time domain of the warped RC pulse is represented as:

\begin{equation}
  RC(t)=\begin{cases}
\dfrac{\pi}{4}  \sinc (\dfrac{1}{2\alpha}), & \text{$\omega(t)= \pm \dfrac{1}{2 \alpha}$}\\
\sinc(\omega(t)) \dfrac{\cos(\dfrac{1}{2\alpha})}{1- (2 \alpha \omega(t))^2} & \text{otherwise} ,
\end{cases}       
\label{eq:RCt}
\end{equation}
while according to Eq. (7), the frequency domain representation $\mathbb{RC}(f)$ is formulated as:
\begin{equation}
\mathbb{RC}(f) =  \int_{\mathbb{R}} R(\nu)  \int_{\mathbb{R}} \sqrt{\dot w (t)} ~ e^{-j 2 \pi( ft-\nu w(t))} ~dt ~ d\nu~,
\label{eq:RCF}
\end{equation}
where 
\begingroup
\small
\begin{equation}
  R(\nu)=\begin{cases}
1, & |\nu| \leq \dfrac{1-\alpha}{2}\\
\dfrac{1}{2} \big(1+ \cos(\dfrac{\pi}{\alpha} [|\nu|-\dfrac{1-\alpha}{2} ] )\big), & \dfrac{1-\alpha}{2} < |\nu| \leq \dfrac{1+\alpha}{2}\\
0 & \text{otherwise}~.
\end{cases}   
\end{equation}
\endgroup

For the constant $\dot{\omega} (t)$ value case, we can observe a RC window in the frequency domain with occupancy proportional to that value. When $\dot{\omega} (t)$ is represented with a sigmoid function, 
\begin{equation}
    \dot{w}(x)=\sigmoid(t/T),
    \label{eq:sigmoid1}
\end{equation}
where the variable $T$ controls the second derivative $\ddot{w}(t)$, i.e., the rate of change of the warping function slope across the $t$ axis. A steep change in the slope of the warping function can be accomplished with a low $T$ value and vice versa.
We can see that the frequency domain window will have spectral components from the warping high slope values and the low slope value. Moreover, the second derivative of the warping function also affects the resultant window shape. If $\ddot{\omega}$ is higher (low $T$), the window occupancy is higher. This is because the frequency domain representation is equivalent to a convolution between the unwarped frequency domain window and the chirp function modulated by $\omega(t)$ as in Eq. (\ref{eq:RCF}). As the chirp function has more occupancy and leaks in the frequency domain \cite{cook2012radar}, the convolution result will have a small leakage outside the expected spectrum.

\subsubsection{Concept 2: The effect of pulse position within the warped time-frequency space}
\begin{figure}[ht]
\centering

    \begin{subfigure}{0.26\textwidth}
    \includegraphics[width=1\linewidth]{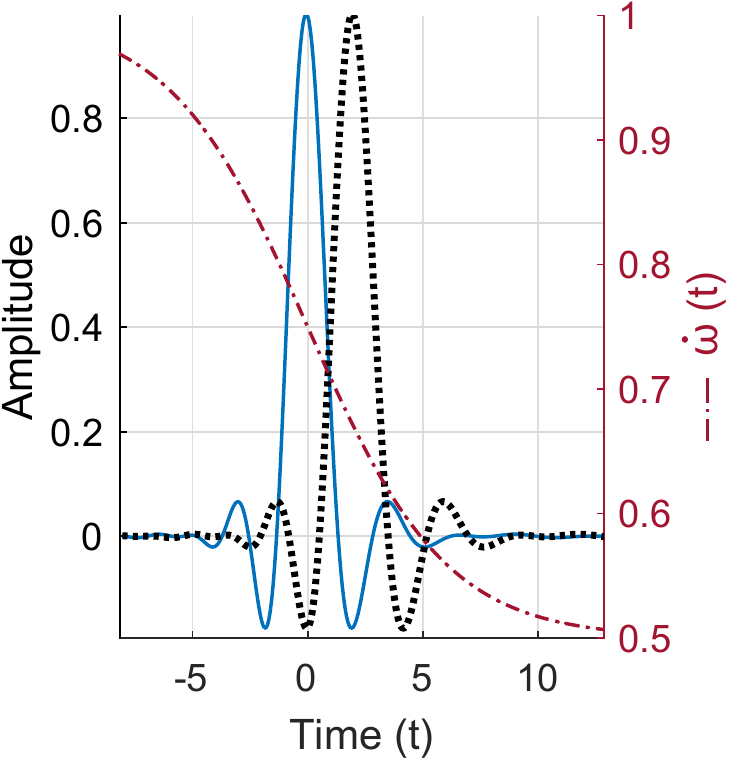} 
 %   \caption{Warping function slope.}
    \label{fig:figBT}
    \end{subfigure}
\begin{subfigure}{0.22\textwidth}
\includegraphics[width=1\linewidth]{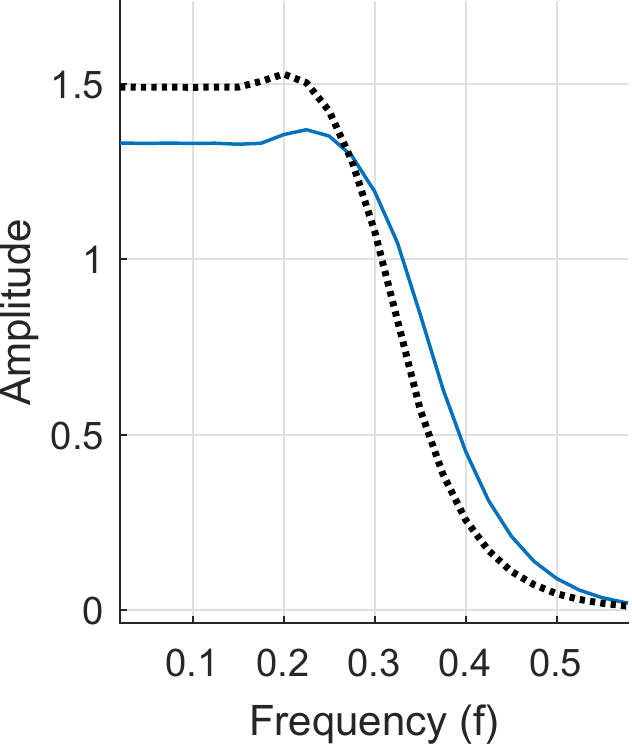} 
%\caption{Freq domain pulse.}
\label{fig:figBft}
\end{subfigure}

\caption{Warping concept 2: One warping profile, different pulse centers.}
\label{fig:figB}
\end{figure}

Now, consider the same warping function with the sigmoid shape shown in Fig. \ref{fig:figB}. The pulses shown in the figure have the same roll-off factor $\alpha=0.33$ but have different positions. The pulse on the left is warped with higher values of $\dot{\omega}$ than the pulse on the right for the same warping function. Due to this, the frequency domain occupancy of the left pulse is higher than the pulse on the right. Therefore, the designed warping function will have different effects on adjacent pulses even if they have the same roll-off factors.  

\subsubsection{Concept 3: Pulse Shapes}

We propose a signal based on zero tail DFT-s-OFDM \cite{Berardinelli}, but with different pulse shaping. Due to spreading with the DFT operation,  ZT-DFT-s-OFDM uses \textit{sinc} shaped pulses. In our proposed pulses, we aim to have a pulse with adjustable tail power (time-frequency occupancy). Raised-cosine pulses are a good candidate, as roll-off factors control the pulse shapes. Also, different RC pulses with different roll-off factors can coexist orthogonally. It was proposed in \cite{Ibrahim,ztdftsofdm} to use pulses with high roll-off factors (low power tails) at the edges of the symbol and pulses with low roll-off factors (high power tails) in the inner part of the symbol. Here, we propose using the same concept with a different pulse shape.

 In this study, we refer to the conventionally raised cosine pulse as a  "symmetric" RC pulse, because later, we define an asymmetric RC pulse. The symmetric RC pulse has an adjustable transition band roll-off parameter $\alpha_n$, and unlike $sinc$ functions, the side lobes amplitude can be adjusted, and the transition band’s rate of decay can be controlled. The symmetric RC pulse shaping function in the time domain is expressed as in Eq. (\ref{eq:RCt}).

Next, we define and motivate asymmetrically shaped RC pulse $P$. The warping modifier compensates for the spectrum expansion caused by applying roll-off factors. Due to that, the warping function slope progresses smoothly over several pulses; we will have different slopes on the two sides of the time domain RC pulse. Hence, it is reasonable to have different roll-off factors for the two sides of the pulse. 
 The lower $\alpha$ is assigned to the side towards the inner part of the waveform (higher $\dot{w}(t)$), and the higher $\alpha$ will be assigned towards the edge side of the waveform (lower $\dot{w}(t)$). This pulse shape results in higher power containment in the frequency domain.

\begin{figure}[ht]
\centering

\begin{subfigure}{0.35\textwidth}
\centering
\includegraphics[width=1\linewidth]{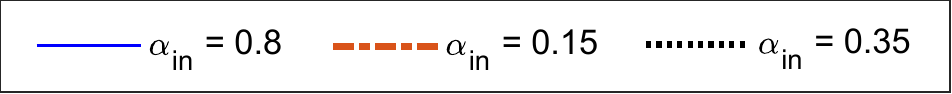} 
\label{fig:Clegend}
\end{subfigure}

    \begin{subfigure}{0.24\textwidth}
    \includegraphics[width=1\linewidth]{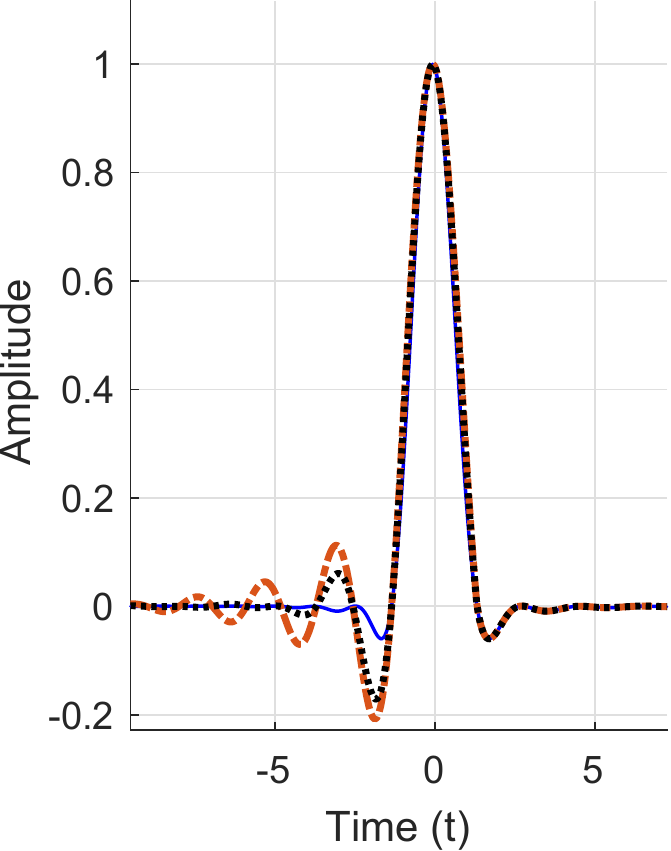} 
 %   \caption{Warping function slope.}
    \label{fig:figCT}
    \end{subfigure}
\begin{subfigure}{0.24\textwidth}
\includegraphics[width=1\linewidth]{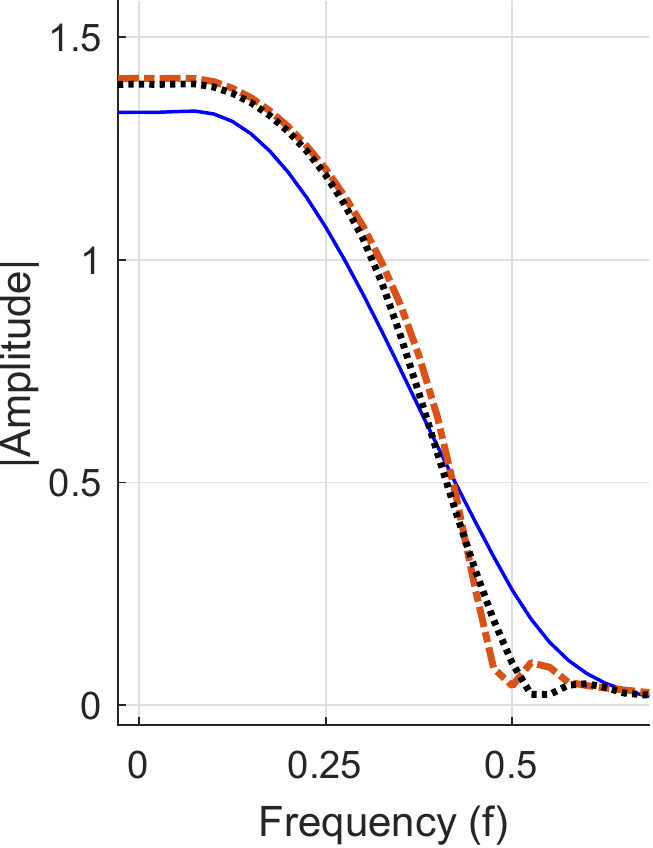} 
%\caption{Freq domain pulse.}
\label{fig:figCft}
\end{subfigure}

\caption{Warping Concept 3: Symmetric versus asymmetric pulse shapes.}
\label{fig:assym_pulse}
\end{figure}
%T=6 
%S_out=.5
%S_in=.99
%ty=cumsum(y)/fs;
% y=(s_out-s_in)*.5*tanh((t-t0)/T)+((s_out+s_in)*.5);

A symbol $S$ of length $L$ pulses, and modulated with $N$ data symbols $a[n]$ as follows:
\begin{equation}
    S(x)=\sum_{n}^N a[n] ~ P_n (\Tilde x-z_h-n),
    \label{eq:ptrain}
\end{equation}
where $n$ is the number of the data symbol, and $z_h$ is the number of zero head symbols. The zero-tails are the rest of the unmodulated symbols ($z_t=L-N-z_h$). 
 {The warping transform is unitary, hence, it is accompanied by an amplitude-changing factor, $\sqrt(w(t))$, as in Eq.} \ref{eq:sqrt}.
 { However, if this factor is incorporated into the proposed modulation, the warping effect would also appear in the constellation of the modulated data. It would result in varying amplitudes of the warped pulses and hence different I/Q constellation representations. To mitigate this effect, we maintain the same amplitude for all the warped pulses, which violates one of the unitary conditions of the warping transform by altering the energy of the pulses but not their orthogonality.}

The asymmetric RC pulse $P_n$ is formulated as follows:  

\begin{equation}
    P_n(x) =
  \begin{cases}
    \frac{\pi}{4 } \sinc(\frac{1}{2 \alpha_{1,n}}) &; x=-\frac{1}{2 \alpha_{1,n}} \\
    \frac{\pi}{4 } \sinc(\frac{1}{2 \alpha_{2,n}}) &; x=\frac{1}{2 \alpha_{2,n}} \\
     \sinc(x) \frac{cos(\pi x \alpha_{1,n} )}{1-(2 x \alpha_{1,n}  )^2} &; x<0 \\
    \sinc(x) \frac{cos(\pi x \alpha_{2,n} )}{1-(2 x \alpha_{2,n} )^2} &; x\geq0 ~,
  \end{cases}
  \label{eq:asympulse}
\end{equation}
where $\alpha_{1,n}$ is the roll-off factor of the left (negative) side of the $n$’s pulse and $\alpha_{2,n}$ is the roll-off factor of the right (positive) side of the $n$’s pulse. 
%\begin{figure}[ht]
%    \includegraphics[width=.45\textwidth]{nonsymetricRC.pdf}
%    %\begin{center}
%    \caption{Non-symmetric warped RC pulse ($\alpha_1=0.7 $, $\alpha_2=0.1 $).}
%    \label{fig:pulse}
%\end{figure}

%\begin{figure}[ht]
%    \includegraphics[width=.45\textwidth]{alphaprofiles.pdf}
%    %\begin{center}
%    \caption{Roll-off factors profiles.}
%    \label{fig:alphaprofiles}
%\end{figure}

Fig. \ref{fig:assym_pulse} shows an example of our used warped pulse shape. The higher $\alpha=0.8$, and the inner $\alpha$ values are $\{0.8, 0.35, 0.15 \}$. The three pulses share the same warping profile, which has a lower slope on the left. We can see that the spectrum shape of the two non-symmetric pulses has lower leakage than the conventional symmetric one. 

The leakage value cannot be solved analytically for such a shape \cite{cook2012radar}. So, in the following analysis, numerical calculations show how different inner roll-off factors can have different leakages. 
 The three pulses share the same warping profile Eq. (\ref{eq:sigmoid1}).
%\begin{equation}
%    \dot{w}(x)=\sigmoid(t/T),
%\end{equation}
%where the variable T controls the second derivative $\ddot{w}(t)$, i.e., the rate of change of the warping function slope across the $t$ axis. 
The warping function has a lower slope on the left $(t<0)$. The spectrum leakage values cannot be solved analytically; therefore, next, we show the impact of changing the inner roll-off factors numerically. 
% We can observe that the pulse has high $\alpha$, hence lower tails power at the left of the pulse, and the opposite is happening toward the inner part of the symbol. 
\subsubsection{Concept 4}
We have the asymmetrically defined RC pulses in Eq. (\ref{eq:asympulse}) with $\alpha_1 = 1$. And we want to find the impact of $\alpha_2$ on the spectral leakage for a warping function that changes its slope from low at the $\alpha_1$ region to high at the $\alpha_2$ region.
%with the formula:

 We sweep the values of $\alpha_2$ at different $T$ values and calculate the leakage ratio $\mathbb{L}$ outside the range $[-0.5, 0.5]$. 
\begin{equation}
\mathbb{L} = 1- \dfrac{\int_{-0.5}^{0.5} \int_{-\infty}^{\infty} P_{\alpha_1 , \alpha_2}(w(t)) e^{-j2 \pi f t} dt df } {\int_{-\infty}^{\infty} \int_{-\infty}^{\infty} P_{\alpha_1 , \alpha_2}(w(t)) e^{-j2 \pi f t}~ dt df } .
\end{equation}

\begin{figure}[h]
\centering
    \includegraphics[width=.45\textwidth]{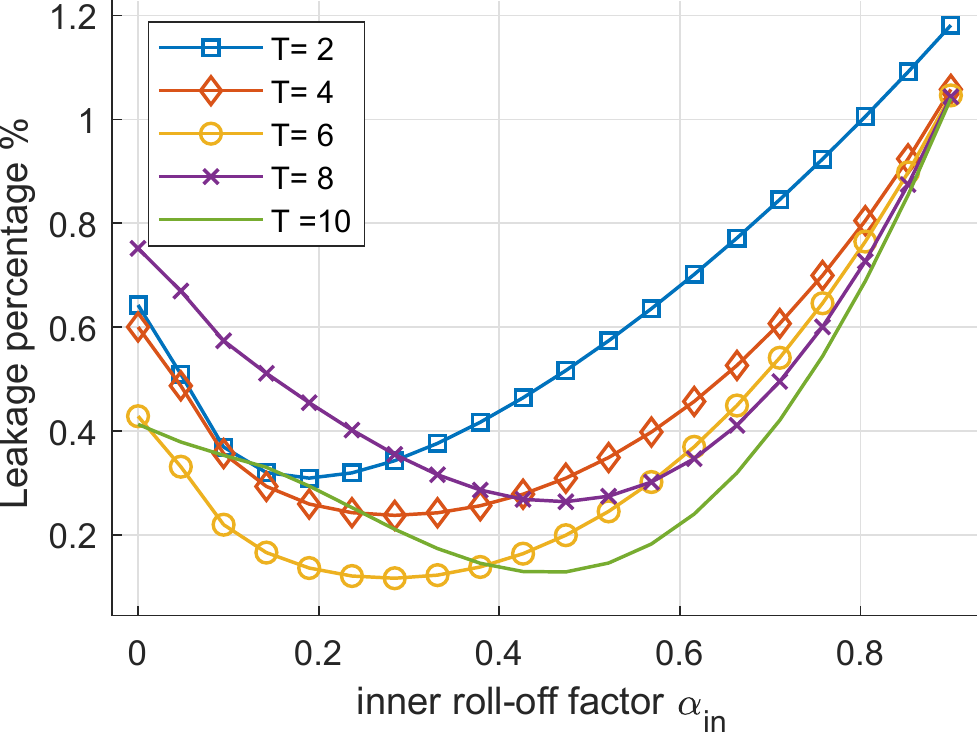}
    %\begin{center}
    \caption{Leakage versus inner roll-off factors.}
    \label{fig:Lvsrolloff}
\end{figure}
We observe from the results in Fig. \ref{fig:Lvsrolloff} that the value of the optimum $\alpha_2$ changes with the warping profile. However, there is no obvious trend in the relationship between the inner $\alpha$, minimum leakage point, and the warping slope represented in $T$. The reason for that is that the leakage value is determined from the Fourier transform of a warped asymmetric pulse, which involves complex constructive and destructive harmonic interferences in the transform domain tails.  
This makes it a nontrivial task to find a closed form for this trend. Therefore the lowest leakage achieved at different inner roll-off factors for different $T$ values needs to be determined with an optimization algorithm. 

\subsubsection{ {Concept 5}}
 {
The warped pulses and the warping function in a sampled system are defined at specific instances in the time domain. The warping function shifts the peak of the sinc or the RC pulse to a different position, resulting in the data or information symbol being moved to a new position in time. It will be demonstrated in the subsequent sections that the warping function should be designed in such a way that the new position precisely aligns with sampling instances.  Therefore, upsampling is necessary to achieve this goal. 
}
\begin{figure}[ht]
\centering
    \begin{subfigure}{0.24\textwidth}
    \includegraphics[width=1\linewidth]{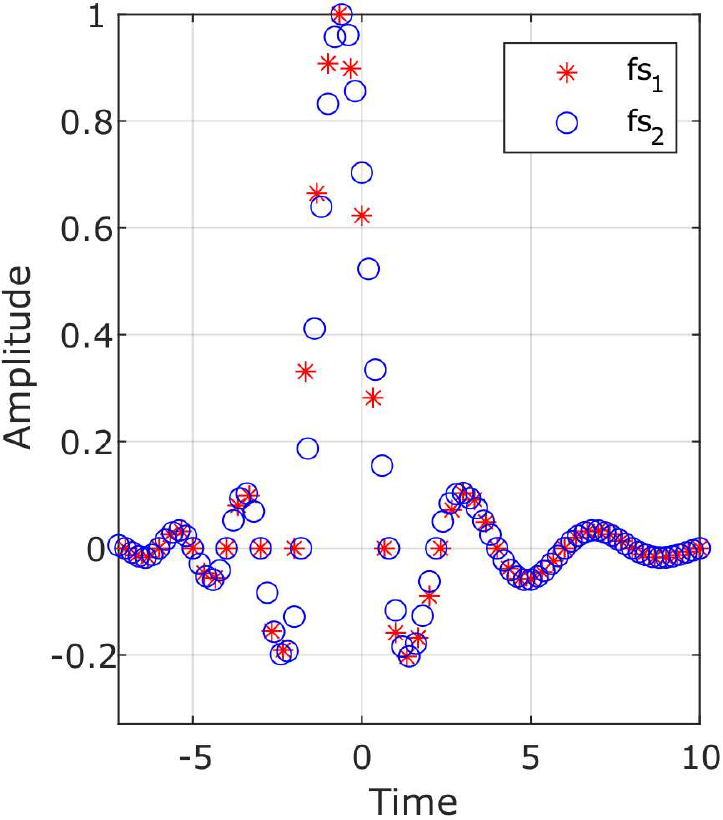} 
 %   \caption{Warping function slope.}
    \label{fig:sampldwarping_time}
    \end{subfigure}
\begin{subfigure}{0.24\textwidth}
\includegraphics[width=1\linewidth]{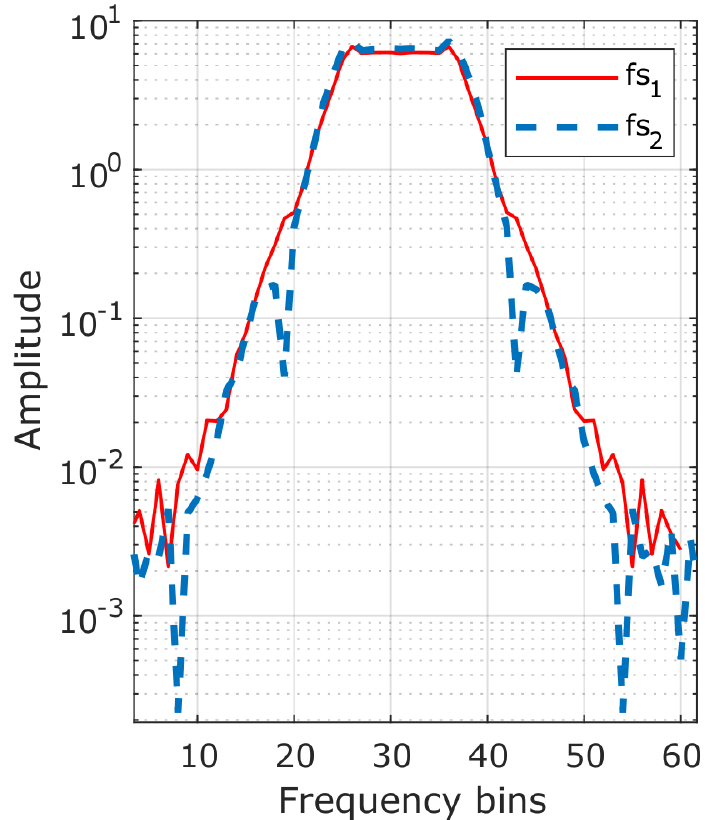} 
%\caption{Freq domain pulse.}
\label{fig:sampldwarping_freq}
\end{subfigure}

\caption{ {Warping Concept 5: The effect of the oversampling ratio.}}
\label{fig:oversampling_ratio}
\end{figure}
 {
The upsampling ratio plays a crucial role in achieving a smooth transition in the warped pulses and reducing the side lobes in the transform domain. To illustrate this concept, Fig. }\ref{fig:oversampling_ratio},  { shows two warped pulses with the same position and warping function but different oversampling ratios.

The upsampling ratio plays a crucial role in achieving a smooth transition in the warped pulses and reducing the side lobes in the transform domain. To illustrate this concept, Fig. \ref{fig:oversampling_ratio} shows two warped pulses with the same position and warping function but different oversampling ratios. 

The pulse with a higher sampling rate $fs_2$ has six samples before and seven samples after the peak, while for the lower rate $fs_1$, there are three samples before and after the peak, with the fourth sample that counts for the warping effect coming later in time. That is per 1T duration of the $RC(t/T)$ function. This lack of a smooth transition leads to unwanted side effects in the frequency domain, resulting in a higher frequency side-band leakage for the lower sampling rate.
Another solution for higher leakage at $fs_1$ is to use a warping function that changes its slope at a lower rate to be more suitable for the low sampling rate.  
The following section includes the waveform design optimization problem.}

\subsection{Warping Function Determination}
This section presents an analysis of designing the warping function based on the prior design of the roll-off factors profile. The roll-off factor in Section III has higher values at the waveform edges and lower values during the inner part. As a result, the high $\alpha$ valued pulses will occupy more space in the transform domain (frequency domain). The warping function is designed to compensate for the high roll-off factors induced expansion. Hence, the first derivative of the warping function $\dot{w}(t)$ should progress from a low value at the edges to a higher value at the middle. 

The warping function slope transitions smoothly from a high value in the middle to a low value at the edges of the symbol. To have this smooth transition between two values, we assume using sigmoid functions for this study. However, other functions can be used, such as exponential functions or higher-order polynomials, as long as smooth transition and monotonicity are preserved.
Our warping function is the addition of two sigmoid functions:
\begin{equation}
\small
\dot{w}(t)=\Bigg(\frac{1}{2}+\frac{1}{2}\sigmoid\Big(\frac{t-t_1}{T} \Big)\Bigg) +\Bigg(\frac{1}{2}-\frac{1}{2}\sigmoid\Big(\frac{t-t_2}{T} \Big)\Bigg).
\label{eq:wdot_sigmoid}
\end{equation}

Then by using the tanh function as the sigmoid function.
\begin{equation}
\small
\dot{w}(t)=\Bigg(- \tanh\Big(\dfrac{t-t_1}{T} \Big) +\tanh\Big(\dfrac{t-t_2}{T} \Big)\Bigg)\dfrac{s_{in}-s_{out}}{2} + s_{out}.
\label{eq:wdot_tanh}
\end{equation}

The parameters that shapes the warping function are 1) edge slopes $s_{out}$ 2) inner slope $s_{in}$ 3) sigmoid centers $t_1$, $t_2$ 4) sigmoid progression $T$.
The shape of the warped RC pulse in the frequency domain is in Eq. (\ref{eq:RCF}). 
%From (7), the shape of the warped RC pulse in the frequency domain is
%\begin{equation}
%\int_{-\infty}^{\infty} RC_{\alpha_n}(w(t)) e^{-j2 \pi f t} dt = 
%\end{equation}
We define the ratio of the spectrum leaked power out of the boundaries $[-f_m , f_m]$
\begin{equation}
\mathbb{L}_n = 1- \dfrac{\int_{-f_m}^{fm} \int_{-\infty}^{\infty} \mathbb{RC}_{\alpha_n}(w(t)) e^{-j2 \pi f t} dt df }{\int_{-\infty}^{\infty} \int_{-\infty}^{\infty} \mathbb{RC}_{\alpha_n}(w(t)) e^{-j2 \pi f t} dt df }.
\end{equation} 

We use linear programming to get the maximum spectral efficiency with the minimum out-of-spectrum leakage \cite{chvatal1983linear}. Linear programming is a technique for optimizing a linear function in order to achieve the best possible result. This linear function or objective function is constrained by linear equality and inequality. The best result is obtained by minimizing or maximizing the objective function. We use the conventional simplex method \cite{nelder1965simplex} to solve the linear programming optimization: 
\begin{equation}
\begin{aligned}
\max_{s_{out},s_{in},t_1,t_2,T} \quad & \Bigg(- \tanh\Big(\dfrac{t-t_1}{T} \Big) +\tanh\Big(\dfrac{t-t_2}{T} \Big)\Bigg). \\ \quad & \dfrac{s_{in}-s_{out}}{2} + s_{out}\\
\textrm{s.t.} \quad & \max_n L < \xi ~, ~~~~
 L = \{ \mathbb{L}_n \}_{n=1}^N , 
\end{aligned}
\label{eq:linprog}
\end{equation}  
where $\xi$, the leakage bound per pulse, has a value that approaches zero. Therefore, the condition ($\max_{n} L <\xi$) ensures that the spectrum is bounded within the ratio limits $\xi$.

\begin{table}[h]
    \centering
    \begin{tabular}{|c c c c c c c|}
    \hline
         & $\alpha_{1,out}$ & $\alpha_{2,out}$ & $\alpha_{3,out}$ & $\alpha_{4,out}$ & $\alpha_{5,out}$ & $\alpha_{6,out}$  \\ 
         \hline\hline
         & 1 & 0.48 & 0.34 & 0.27 & 0.17 & 0.08   \\
         \hline
    \end{tabular}
    \caption{Roll-off factors of the pulses from outside to inside.}
    \label{tab:problem}
\end{table}

\begin{table}[h]
\begin{center}
\begin{tabular}{||c c c c c c||} 
 \hline
    & $S_{in}$&$S_{out}$&$t_{edge}$&  &$T$\\ [0.2ex] 
 \hline\hline
  Sym. P & 0.98 & 0.49 &5.3 &-5.3 & 1.8 \\ 
 \hline
   Asym. P & 0.98 & 0.49 &5.7 & -5.7 & 3\\
 \hline\hline
   $\alpha_{1,in}$&$\alpha_{2,in}$&$\alpha_{3,in}$&$\alpha_{4,in}$&L & D \\ [0.2ex] 
 \hline\hline
 1 & 0.48 & 0.34 & 0.27 & 0.02\% & 14.06\\ 
 \hline
 0.3 & 0.1 & 0.08 & 0.08 & 0.02\% & 13\\
 \hline
\end{tabular}
\caption{Symmetric versus Asymmetric pulses solution 1  L=0.3\%.}
\label{tab:solution}
\end{center}
\end{table}

In the case of asymmetric pulses, extra parameters will be added to the optimization problem. The inner $\alpha$ values of the edge pulses have optimum values that change with the warping profile, changing the optimization boundaries and allowing for higher spectral efficiency. The rolls-off factors profile generated from Eq. (\ref{eq:case3}) is shown in Table \ref{tab:problem}, for six pulses. This profile is mirrored and used for the next six pulses. In Table \ref{tab:solution} we show a solution for a 12-pulses waveform with RC and asymmetric RC pulses. The value $D$ represents the warping expansion in the time domain from the first pulse to the last pulse:
\begin{equation}
  D= {w}^{-1}(-5.5)-{w}^{-1}(5.5).
\end{equation}

\begin{figure}[h]
    \centering
    \includegraphics[width=.45\textwidth]{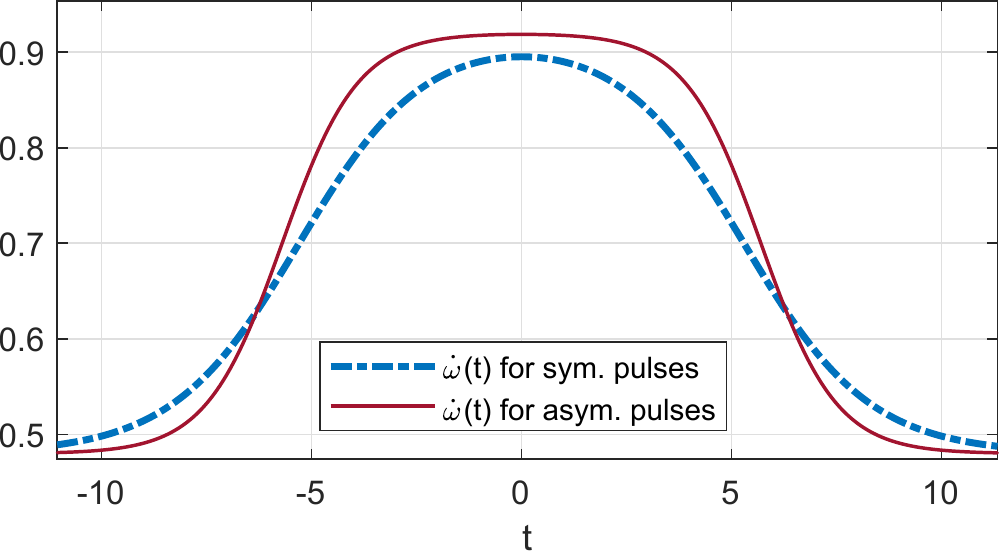}
    \caption{The slopes of the warping function for the 12-pulses problem, for the symmetric and asymmetric pulse shapes.}
    \label{fig:symvsasym_w}
\end{figure}

In the absence of any warping, for 12 pulses, this value will be $11T$. By substituting Eq. (\ref{eq:wdot_sigmoid}) we plot the slope of the warping profile for the two solutions as shown in Fig. \ref{fig:symvsasym_w}.

To perform the warping with no interpolation errors, the sampling duration $T_s$ should be less than the data symbol duration. The time samples $mT_s$ should be mapped to the integer $n$ in the warped domain, where $m$ is an integer, and $n$ is the number of the pulse in the pulse stream. Otherwise, there will be sampling offsets, and inter-symbol (pulse) interference will occur. We propose to fit the above warping function to a piecewise spline warping function to maintain orthogonality between pulses.

\begin{figure}[t]
\begin{center}
    \includegraphics[width=.48\textwidth]{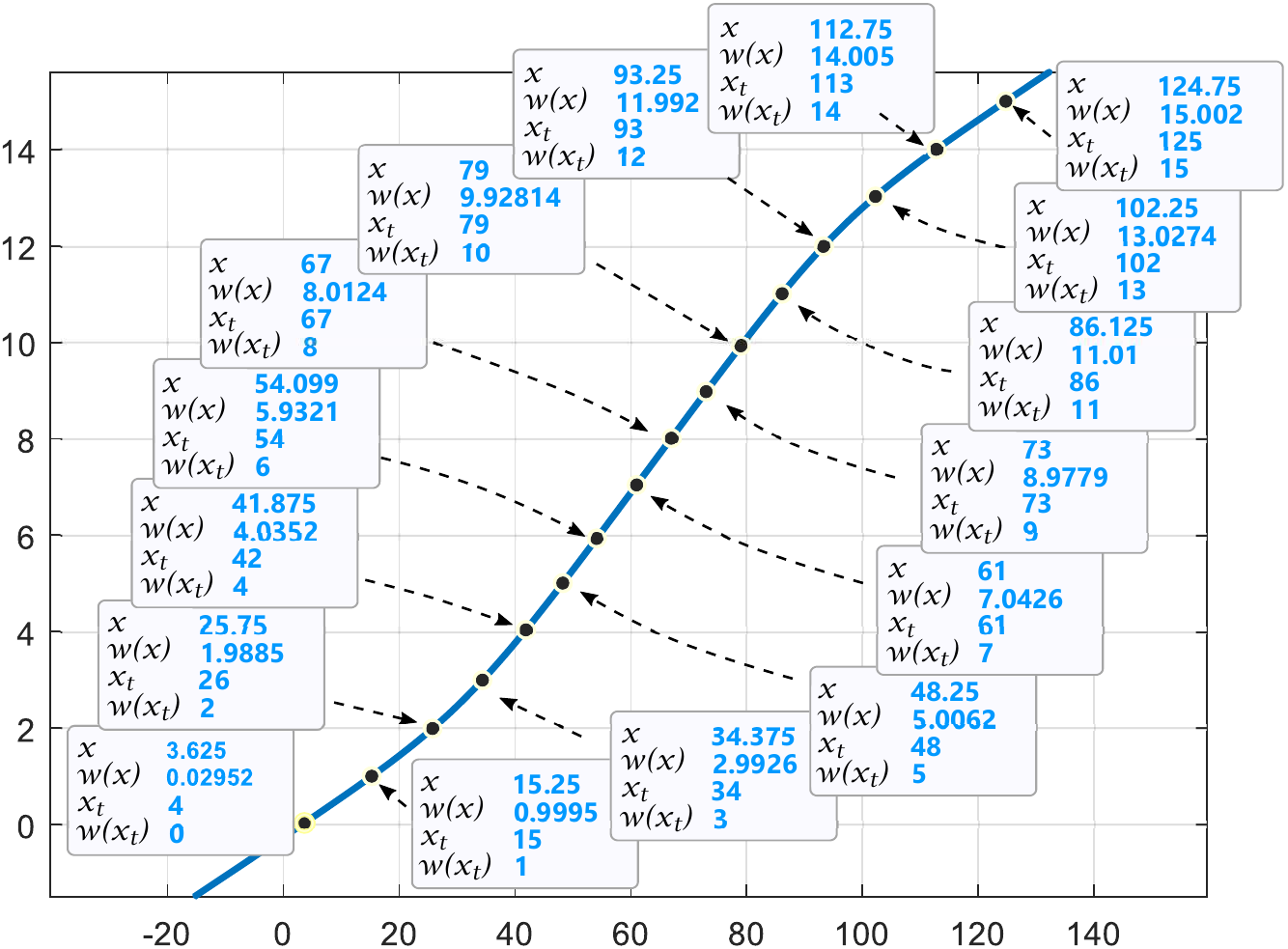}
    \caption{Warping function interpolation for 12 pulses.}
    \label{fig:warpingfuncinterp}
    \end{center}
\end{figure}

The spline interpolator uses a piece-wise polynomial to fit small subsets of the interpolated function between specific values called the knot samples. The spline segments are low-degree polynomials which makes it better than using a single high-degree polynomial that fits all the points. For $n$ pulses, we need $n+1$ polynomial segments and $n$ knots, and the simplest polynomials are the cubic splines with the conditions:   
\begin{equation}
\begin{aligned}
    & q'_i(x_i)=q'_{i+1}(x_i) \\
    & q''_i(x_i)=q''_{i+1}(x_i),
    \end{aligned}
    \label{eq:knotconditions}
\end{equation}
where $q'_i$ and $q''_i$ are the first and second derivative of the segment $q_i$ at the knot $x_i$.
In our case, the knot samples are the points where the pulses exist on the oversampled axis. As shown in the example in Fig. \ref{fig:warpingfuncinterp}, the new spline warping function maps the discrete time domain samples $x_t=w^{-1}(n)$ to the nearest integer values and the warped values $w( w^{-1}(n) )$ to the nearest $n$. 
 In this example, a length of 128 samples is intended to represent the warped symbol, which will require an inverse Fast Fourier transform (IFFT) length of 128 at the receiver, as will be discussed in the following section. 

%Another aspect of the warping function is that it is used to dilate the edge pulses with higher roll-off factors to occupy the same spectrum occupied by the inner pulses. Therefore, for the warping function to be compatible with the proposed waveform, the slope of $w(x)$ should be lower at the edges than at the middle of the symbol. The change in the slope is gradual to follow the gradual change of the roll-off factors.

\section{Transceiver Design}
This section proposes a transceiver design for the presented time-frequency warped waveform. We show the waveform representation along the transceiver chain, along with the transceiver blocks needed to accomplish the modulation and demodulation.
The sampled warping function at the sampling instances, $(w(x_t) ~; ~x_t \in \mathbb{Z})$, should have integer values at the sampling points $w^{-1}(n)$. Otherwise, interpulse interference will occur as the peaks and the zeros of the pulses will fall in between sampling points. Therefore, the first derivative of the warping function $\dot{w}(x)$ is quantized to match the above condition. And the warping function becomes a piecewise function that follows a sigmoid profile. We show in Fig. \ref{fig:txrx} the signal representation at different stages of the transceiver chain.
\subsection{Transmitter}

\begin{figure}[ht]
    \begin{center}
    \includegraphics[width=.32\textwidth]{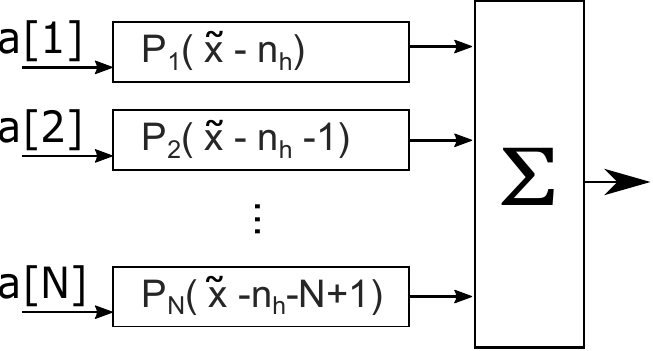}
    \caption{Proposed transmitter block diagram.}
    \label{fig:txscheme}
    \end{center}
\end{figure}
\begin{figure}[h]
     \centering
     \begin{subfigure}[b]{0.45\textwidth}
         \centering
         \includegraphics[width=\textwidth]{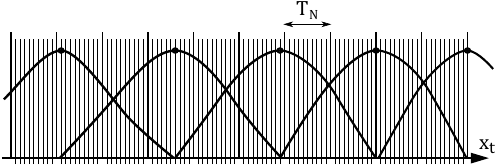}
         \caption{}
         \label{fig:txrx_a}
     \end{subfigure}
     \begin{subfigure}[b]{0.45\textwidth}
         \centering
         \includegraphics[width=\textwidth]{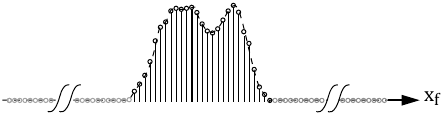}
         \caption{}
         \label{fig:txrx_c}
     \end{subfigure}
     \begin{subfigure}[b]{0.25\textwidth}
         \centering
         \includegraphics[width=\textwidth]{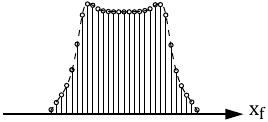}
         \caption{}
         \label{fig:txrx_d}
     \end{subfigure}
     \begin{subfigure}[b]{0.4\textwidth}
         \centering
         \includegraphics[width=\textwidth]{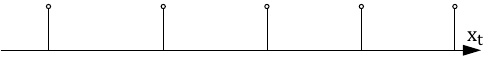}
         \caption{}
         \label{fig:txrx_e}
     \end{subfigure}
    \caption{Wavefrom stages along the transceiver chain.}
    \label{fig:txrx}
\end{figure}

The transmitter is implemented using a filter bank scheme, as shown in Fig. \ref{fig:txscheme} because there are several pulse shapes included in the same waveform. The complex data symbols $a[n]$ are modulated by the pulse shapes at each branch then the results are added using a summation block, satisfying the modulation in Eq. (\ref{eq:ptrain}) and Eq. (\ref{eq:asympulse}). Each branch is shaped with the corresponding pulse $P_n(\Tilde{x} - {n} - z_h)$, by considering the spacing of zero heads and zero tails. The sampling rate of the filter is higher than the data symbol rate to satisfy the warping function's characteristics. 
As mentioned above, to define the warped axis $\tilde{x}=w(x_t)$, a piecewise cubic spline is used. 
\begin{equation}
    \Tilde{x} = 
    \begin{cases}
        a_0 x_t^3 + b_0 x_t^2 + c_0 x_t + d_0 & ; 0< w^{-1}(x_t) \leq 1 \\
        a_1 x_t^3 + b_1 x_t^2 + c_1 x_t + d_1 & ; 1< w^{-1}(x_t) \leq 2 \\
        ~~~~~~~~~~~~~~~~\vdots & \\
        a_n x_t^3 + b_n x_t^2 + c_n x_t + d_n & ; n< w^{-1}(x_t) \leq n+1 \\
        ~~~~~~~~~~~~~~~~\vdots & 
    \end{cases},
    \label{eq:splinewarp}
\end{equation}
where the piecewise intervals are between the spline knots while the parameters, $a_n,~ b_n,~ c_n,$ and $d_n$ satisfy the conditions in Eq. (\ref{eq:knotconditions}),

We can observe the shape of the individual waveform pulses and their positions in the oversampled time domain in Fig. \ref{fig:txrx_a}. $T_N$ represents the duration of the Nyquist rate. 

%In  Fig. \ref{fig:txrx_b}, we show the frequency domain of the waveform. Note that the window is not rectangular shaped; instead, it has the warping effect as a convolution with a chirp spectrum. 
%Also, the empty band corresponds to the oversampling effect.

\subsection{Receiver}
The receiver scheme used in our work is identical to the DFT-s-OFDM receiver but with a higher sampling rate to account for the nonuniform spacing between the symbols. The reception window is rectangular-shaped, as defined by the FFT operation, which means that the receiver filter is unmatched with the raised cosine-shaped filter of the transmission pulses. However, the received symbols are free of ISI. The ISI-free property holds because the transmitted pulses are designed to be orthogonal (ISI-free) \cite{634674,705392}. The reason that the reception filter is unmatched is that the transmitted symbol has several roll-off factors, hence the difficulty of being matched to all of them.   

\begin{figure}[ht]
    \begin{center}
    \includegraphics[width=.48\textwidth]{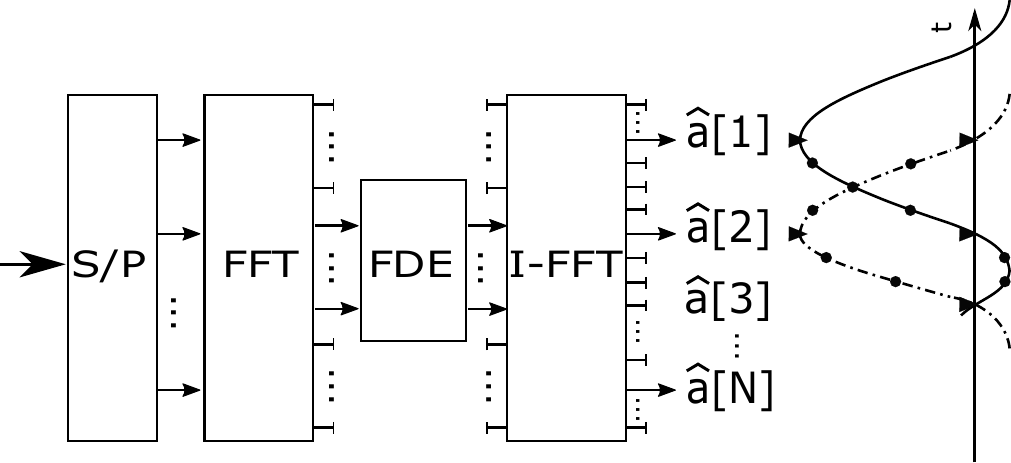}
    \caption{ {Proposed receiver block diagram.} \textsuperscript{1}  }
    \small\textsuperscript{1} The FFT and IFFT blocks have an "FFT-shift" operation embedded.
    \label{fig:rxscheme}
    \end{center}
\end{figure}

As shown in Fig. \ref{fig:rxscheme}, after single to parallel conversion, the signal is transformed to the frequency domain using a fast Fourier transform (FFT) block. The signal is then equalized via a frequency domain equalization (FDE) block. Then transformed back to the time domain via an IFFT block. The estimates of the data symbols are then extracted after nonuniform downsampling with the exact nonuniform mapping at the transmitter. 

When the signal is in the air, the multipath channel affects the frequency domain shape; also, the noise will be added, as shown in Fig. \ref{fig:txrx_c}. We keep showing the empty sidebands because the receiver captures the signal with oversampling $V$. The signal is received in the time domain, hence, Fig. \ref{fig:txrx_c} represents the output of the receiver FFT. The DFT operation run by the FFT block results in the received frequency domain representation
\begin{equation}
    R(x_f) = \sum_{x_t=0}^{VN-1} r(x_t). e^{-j \frac{ 2 \pi }{VN} x_f x_t}.
    \label{eq:rx_fft}
\end{equation}

Practically, the sidebands will not be zero; but they contain the neighboring transmissions, and they will be filtered out.
Therefore, the FFT output of the center band is the frequency bins that we will keep, and the other side frequency bins are discarded. Note that, in Fig. \ref{fig:rxscheme}, the FFT block has an "FFT-shift" operation embedded. An FFT-shift operation rearranges the frequency bins to put the zero-frequency bin in the middle of the output array. The same is done for the IFFT block input. 
For an $x_f$ range $[0,~VN-1]$ ,in Eq. (\ref{eq:rx_fft}), we keep the bins $[0,~N-1] \cup [VN-N,~VN-1 ]$.

To reduce the receiver complexity,  FFT pruning \cite{prunedfft} is used for the non-used bins. FFT pruning is a technique to efficiently compute the discrete Fourier transform (DFT) for a subset of input points or output points or a subset of both. Further details are in the following sections.  

\begin{figure*}
    \centering
    \includegraphics{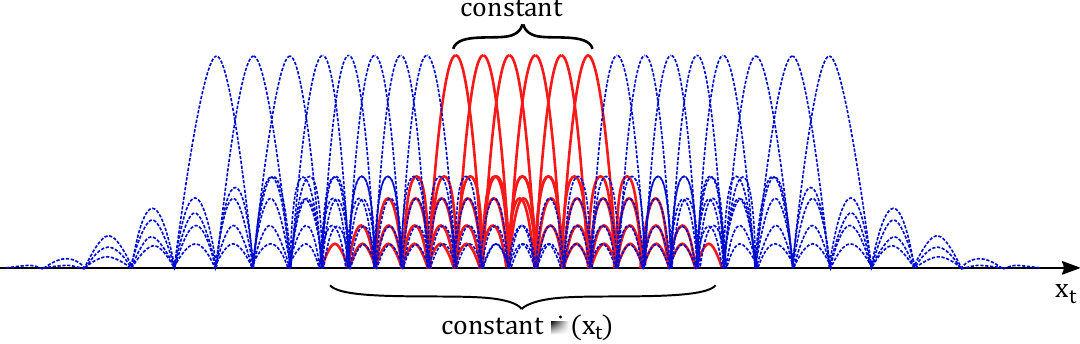}
    \caption{Splitting modulation regions for lower computational complexity.}
    \label{fig:tx1_complxty}
\end{figure*}

FDE can be performed either using minimum mean square error MMSE or zero-forcing equalization, but comparing the equalizers is outside the scope of the paper. For zero-forcing equalization, the resultant vector is 
\begin{equation}
    R_e(x_f)=\dfrac{R(x_f)}{H(x_f)} ~s.t.~x_f =  [0, 2N-1].
    \label{fig:rx_fde}
\end{equation}

 {Additionally, the choice between time domain equalization (TDE) and frequency domain equalization (FDE) depends on the specific characteristics of the wireless channel and the system requirements. Although the DFT-s-OFDM scheme is capable of and compatible with frequency domain equalization, the system can still equalize using the time domain equalization schemes. Nevertheless, the unequal spacing of the warped symbols makes it difficult to space the channel impulse response taps with the symbol rate. However, it is possible to space them with the oversampling rate since it is equispaced. This might entail the usage of a training sequence that differs in its characteristics from our proposed warped waveform. }

The resultant values of the frequency bins are shown in Fig. \ref{fig:txrx_d}. This spectrum has in its content nonuniformly sampled information in the time domain that we want to retrieve. 
A nonuniform IFFT can be used for this task \cite{dutt1993fast, greengard2004accelerating}. However, existing nonuniform FFTs are essentially a combination of a local interpolation scheme and the standard FFT, which would introduce interpolation errors in the resultant vector. 

Instead of interpolation, we will realize the nonuniform samples via high-resolution transform content by pruning \cite{sreenivas1980high}. This enables the extraction of the samples at the pulse positions predefined from the warping function. Further details on the pruning method and its computational complexity are presented next. 
The IDFT operation at the IFFT block retrieves the equalized waveform in the time domain,
\begin{equation}
    r_e(x_t)=\sum_{x_f \in baseband} R_e(x_f). e^{j \frac{ 2 \pi }{VN} x_t x_f},
    \label{eq:rx_ifft}
\end{equation}
where the output of the FDE block is assigned to the $baseband \in [0,N-1] \cup [VN-N, VN-1]$ interval. We are interested in the output samples at $x_t=w^{-1} (n)$. 
 { We show the corresponding positions of those received symbols $\hat{a} [n]$  on the oversampled warped pulses output by the IFFT block, in Fig.} \ref{fig:rxscheme}. 
Therefore, the IFFT block will benefit from pruning at its input and output. Further, the computational complexity reduction of pruning is presented in the next subsection. 

\subsection{Computationally Efficient Methods}

This section proposes methods for efficient computational complexity at the transmitter and the receiver. First, we propose the availability to reduce the complexity of the transmitter filter bank by considering the $\dot{w}(x_t)$ function with a flat middle. Then we give further details on the pruned FFT and IFFT blocks at the receiver. 
\subsubsection{Transmitter}

\begin{figure}[h]
    \centering
    \includegraphics[width=.38\textwidth]{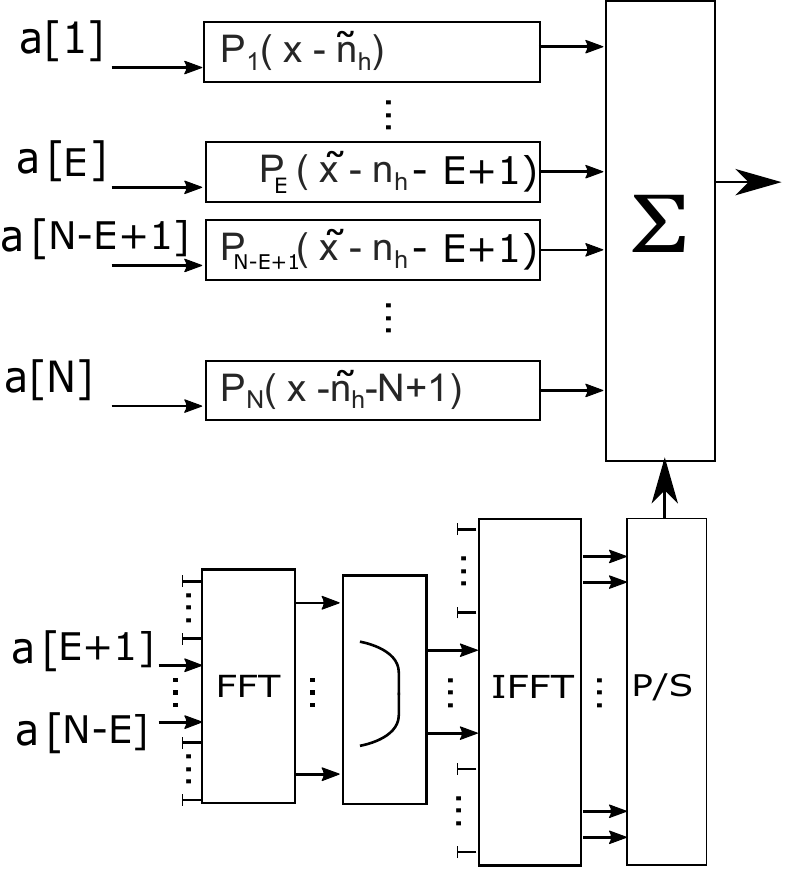}
    \caption{Splitting modulation regions transmitter scheme.}
    \label{fig:tx2_complxty}
\end{figure}

For symbols with a relatively high number of pulses, a middle region of pulses has the lowest $\alpha$ value, which is constant for this region of pulses. The result of this is the availability of a warping function with the same slope over this region, taking into consideration the duration of the effect of these pulses on the time axis. Fig. \ref{fig:tx1_complxty}  shows the middle pulses with the solid red line and the edge pulses in the blue dotted line. The constant $\dot{w}(x_t)$ region extends to include the middle pulses with their tails. This extra duration will depend on the minimum $\alpha$ value. The middle pulses are not warped; therefore, they can be modulated by a parallel structure of the DFT-s-OFDM modulator, as shown in Fig. \ref{fig:tx2_complxty}. The DFT-s-OFDM part has a long IFFT output to match the sampling rate of the filter bank part responsible for the warped pulses. 

The number of filter bank branches is the number of the edge pulses $2 E$. The summation block adds the two streams of the filter bank structure and the DFT-s-OFDM structure, and the edge pulses' corresponding positions are left empty in the DFT-s-OFDM structure. A block between the FFT and the IFFT is responsible for the pulse shaping with the middle $\alpha$ value. The block input and output sizes are as follows: 
\begin{itemize}
    \item The FFT has $N$ inputs and $2N$ outputs to account for the windowing space. 
    \item The windowing block has $2N$ inputs and $2N$ output. 
    \item The IFFT block has $2N$ inputs and $
    V.N$ outputs, where $V$ is the upsampling rate. 
\end{itemize}

 Splitting the modulation regions as proposed reduces the computational complexity due to using the efficient FFT and the IFFT blocks. Furthermore, the IFFT block can be more efficient if pruning is used to exclude $VN-2N$ inputs from the IFFT butterfly structure. We assume a radix-2 decimation in frequency (DIF)-IFFT with Skinner pruning \cite{Skinner} for this block. FFT pruning for $N$ inputs and $L$ outputs has computational complexity $O(NlogL)$. 
The computational complexity of the proposed two-mode transmitter is $O (VN~log~N + VNE)$ instead of $O(VN^2)$ for the full filter bank scheme. 

\begin{figure}
    \centering
\includegraphics[width=.48\textwidth]{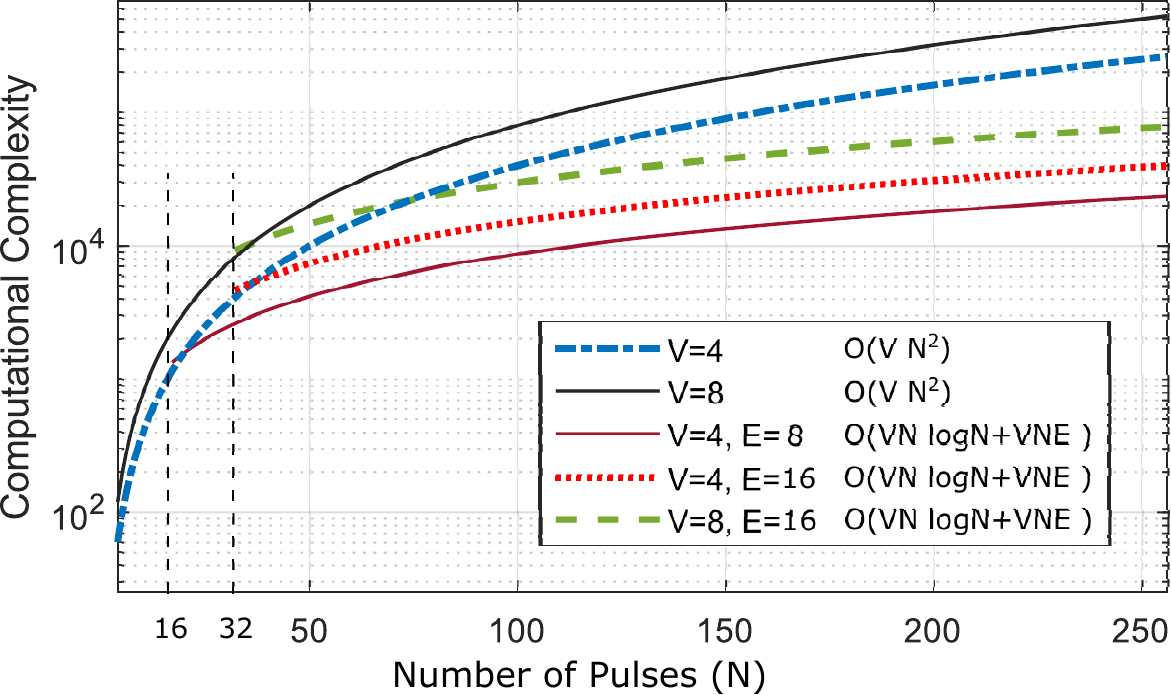}
    \caption{{Transmitter computational complexity.}}
    \label{fig:tx_comp}
\end{figure}

 {The complexity of the transmitter is plotted in Fig.}\ref{fig:tx_comp} { as a function of N for various V and E values.  The upper two curves show a quadratic increase in power, while the lower three curves correspond to the proposed modulation splitting approach. The proposed reduction is valid in the regions $N>2E$, and it appears to branch off from the higher complexity curve with the same V value. This indicates that beyond a certain value of E, determined by the warping profile, it is advantageous to use the proposed modulation splitting technique.}

%The butterfly structure of this block is shown in the Appendix. 

% The multiplication operation is the most expensive operation; hence, we only take then them into consideration for the computational complexity evaluation.  
 %The number of multiplications in the filter bank part is $2E \times VN$. While the number of multiplications in the DFT-s-OFDM modulator part is composed of several terms. The $2N$ FFT block has $N \times log_2 (2N)$ multiplications, the windowing block has $2N$ multiplications, and the pruned $VN$ IFFT block with $2N$ inputs has $2VN\times log_2(2N)$ multiplications \cite{prunedfft}. This sums up to
 %\begin{equation}
 %    MUL_{tx}= 2EVN + N  log_2 (2N) + 2N  +2VN log_2(2N)
 %\end{equation}
 %multiplications, versus $VN^2$ for the full filter bank implementation. In Fig. (), we plot the number of %multiplications.   

\subsubsection{Receiver}
The receiver block in Fig. \ref{fig:rxscheme} has two blocks that can benefit from pruning. The block input and output sizes are as follows: 
\begin{itemize}
    \item The FFT of size $VN$ has $VN$ inputs and $2N$ outputs to account for the windowing space. 
    \item The FDE block has $2N$ inputs and $2N$ output. 
    \item The IFFT of size $VN$ has $2N$ inputs and $
    N$ outputs, where $V$ is the upsampling rate. 
\end{itemize}
The computational complexity reduces to $O(VNlogN/V)$ instead of $O(VN log VN)$ (see Eq. (4) in \cite{sreenivas1980high}, assuming $2 log_2 (2N) > log_2(VN)$). A pruned FFT can either be DIT or DIF type. We propose a radix-2 DIF FFT with Skinner's pruning at the output for the FFT block. For the IFFT block, we propose a radix-2 DIT IFFT with Markel's pruning \cite{markel1971fft} at the input and Skinner's pruning at the output. 
%Diagrams for the mentioned schemes are shown in the Appendix. 

\begin{figure}
    \centering
\includegraphics[width=.48\textwidth]{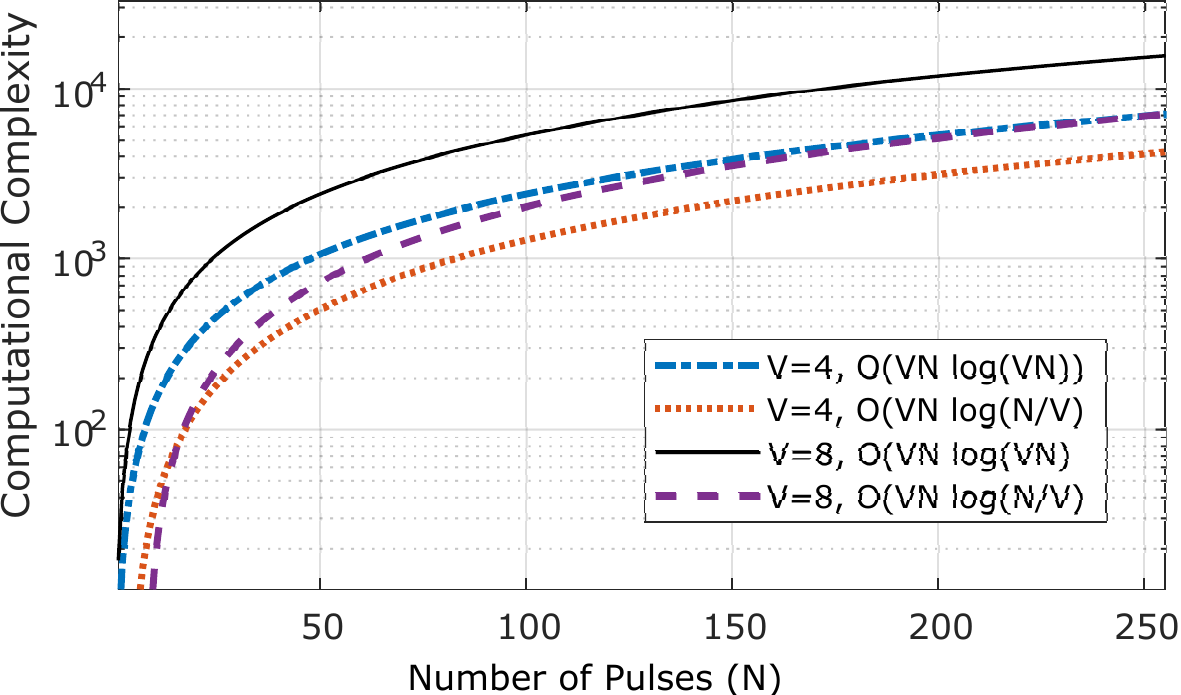}
    \caption{{Receiver computational complexity.}}
    \label{fig:rx_comp}
\end{figure}

 {For the receiver, the pruning reduction is shown in Fig. }\ref{fig:rx_comp} {. We can observe that the complexity increases with V, but the pruning becomes more effective at higher V. }

\subsection{System Level Optimization (RL State Action Space)}
The design space, degrees of freedom, and flexibility are high in the proposed waveform. In this section, we discuss the system parameter trends. Then we introduce a reinforcement learning-based optimization approach to flexibly design the proposed waveform between a transmitter and a receiver.  On an arbitrary link between a transmitter and a receiver, there are a group of optimizable tradeoffs that are dependent on the receiver and transmitter requirements, capabilities, and the communication scenario priorities.

Power consumption $Pw$, spectral efficiency $Se$, out-of-band emission $E_f$, time domain zeros suppression $E_z$, bit error rate $BER$, and tolerance to temporal $\tau_s$ shifts are among the parameters that need to be optimized over a communication link. 
 {A utility function representing the link requirements denoted by $U (Pw ,~ Se ,~ E_f ,~ E_z ,~ BER ,~ \tau_s)$, at specific values of the above parameters, has the gradient $\nabla U$
corresponding to the weight or the importance of each requirement in relation to the others. }
\subsubsection{Parametric Trends}
Here we discuss the relationship between three main parameters and the rest of the system's variables; oversampling rate, roll-off factor profile, and the number of pulses. 

The oversampling enables an interpolation-free receiver for the warped waveform. With higher oversampling, we can have less abrupt transitions for the discrete warping function, hence, lower frequency OOBE. Moreover, this can allow a steeper warping function, which allows for a steeper roll-off factor profile. And a symbol with a steep roll-off factor profile (low $T$ parameter in Eq. (\ref{eq:linprog})) can be well-contained with short bursts of pulses $N$, hence lower latencies. Given that the pulses can carry higher modulation orders, the drawback of oversampling is the increased complexity of the transmitter, the receiver, and the corresponding energy expenses. 

The roll-off factor profile controls the compromise of higher spectral efficiency versus higher containment due to lower time-domain zero tails, given that the warping function conforms to the same spectral window for all the pulses. Low time-domain zero tails are better for temporal shift offsets. But, again, the steep $\alpha$ profile may need a higher $V$.

When low latency and short bursts are not needed, a higher number of pulses can allow higher spectral efficiency. The reason for that is that more inner spectrally efficient pulses can be packed along with the edge lower spectrally efficient pulses, as in Section V.C.1. Therefore, on average, the spectral efficiency of the whole symbol will be higher. On top of that, a relaxed roll-off factor profile and warping function can be used (high $T$ parameter in Eq. (\ref{eq:linprog})). Hence, a lower $V$ can be acceptable. However, increasing $N$ will increase the FFT/IFFT sizes at the receiver and, consequently, the complexity. This means that there will be a sweet spot for the  $N$ versus $V$ depending on the link requirements.

Higher spectral containment and lower zero tails lead to lower interference levels between adjacent symbols in time and frequency domains. This translates into lower BER in the presence of adjacent interferers, as shown in the Simulations Section. The gain in throughput will depend on the SNR level of operation, higher modulation order, and error coding scheme,  Therefore, we propose next the RL-based optimization method that customizes the waveform parameters based on the link needs.

\subsubsection{RL Based Optimization}
In this section, we propose a state-action space for the above optimization problem in a reinforcement learning context. RL is a powerful instrument that can reach policies and methods that go beyond simple human decision-making procedures \cite{kaelbling1996reinforcement, silver2017mastering}. The agents in this paradigm use a comprehensive state-action reward function or table (Q-table), which is a mapping of actions with predicted rewards. Based on this, the agent chooses one of the potential actions. As a result, it transitions to a new state with a different reward. The agent's ultimate purpose is to collect as much cumulative reward as possible.
RL has been used in communication systems for different optimization problems \cite{9043893,8683970,IbrahimComplexty}. In this section, we are not proposing a specific RL learning algorithm, which can be explored in a future study; however, we are limiting our suggestions to the state-action space design.  
% ask dr umair for his papers to cite here
 The RL problem is well described by a Markov Decision Process (MDP). MDP systems obey the Markov property where the transitions of the process depend only on the current state and actions and not the prior history. The MDP is described by the tuple $<S,A,R,T, \gamma >$, where the elements of the tuple are:
\begin{itemize}
    \item $S$ is the set of all valid states that can be discrete or continuous. 
    \item $A$ is the set of all valid actions.
    \item $R: S \times A \times S \mapsto  \mathbb{R}$ is the reward function, with $r_t = R(s_t, a_t, s_{t+1})$, and $t$ is the time step unit. 
    \item $T: S \times A \mapsto T(s')$ is the transition probability to state $s'$ if action $a$ taken at state $s$. 
    \item $\gamma \in [0,~1]$ is the discount rate mapping to the future rewards.
\end{itemize}

% The trial and error process of the agent is supported by exploration, and its purpose is to try states searching for higher rewards. The process of using the knowledge already collected is called exploitation. Between exploration and exploitation the agent tries to find the balance for optimal learning and rewards maximization.

\subsubsection*{Action-Value Functions} 
The action-value function (Q-function) represents the expected return of the state-action pair $(s,a)$, under policy $\pi$. Actions are extracted from the value functions such that we maximize the expected returns.
\begin{equation}
\begin{aligned}
    Q^\pi (s,a) =  E \Big[ \sum_{t \geq 0} ~ & \gamma^t R(s_t, a_t, s_{t+1}) ~|~  a_t \sim \pi( \cdot | s_t) , \\ & s_0=s, a_0 =a \Big].
    \end{aligned}
\end{equation}

The above equation means that the expected return is determined from all of the future state action routes decided by the policy and discounted by the factor $\gamma$. 
The Q-function is updated when transitioning from state-action $(s,a)$ to new state $s'$, and a reward $r$ is received
\begin{equation}
    Q (s,a) \longleftarrow (1-\alpha) Q(s,a) + \alpha [r + \gamma \max_{a'} Q(s',a')],
\end{equation}
where, $\alpha$ is the learning rate. Next state $s'$ is sampled from the environment's transition rules $s' \sim P := s' \sim P(\cdot | s,a)$. The next action is sampled from the policy rules $a' \sim \pi := a' \sim \pi(\cdot | s')$.
%The optimal action-value function $Q^* (s,a )$ gives the expected return if the agent starts at $(s,a)$ and keeps acting according to the optimal policy.
%\begin{equation}
%\begin{aligned}
%    Q^*(s,a) = \max_\pi   E \Big[ \sum_{t \geq 0} ~ & \gamma^t R(s_t, a_t, s_{t+1}) ~|~  a_t =a^* , \\ & s_0=s, a_0 =a \Big].
%\end{aligned}
%\end{equation}

In our learning system, rewards are calculated at the receiver and transmitter, reflecting the system preferences. The Q-function is shaped by the rewards to guide the agents toward the maximum reward. 
%This can become a future direction in studying this topic. 

\begin{figure}[h]
    \centering
    \includegraphics[width=.48\textwidth]{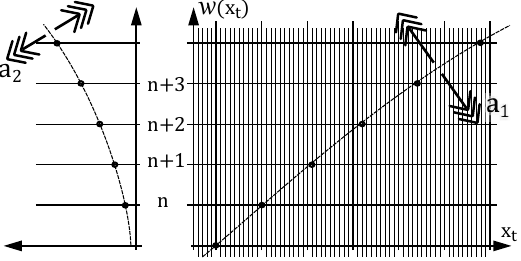}
    \caption{Learning actions shown on warping and roll-off profile segments.}
    \label{fig:rl_warp}
\end{figure}

\subsubsection*{State-Action Space}
For an efficient learning process, it is essential to represent an abstract learning problem with low state-action dimensionality \cite{Li2006TowardsAU}.  
We consider the warping and roll-off profiles in Fig. \ref{fig:rl_warp} with a reduced representation rather than the ones defined above equations, Eq. (\ref{eq:3}), and Eq. (\ref{eq:linprog}). The system designer can choose from several options\cite{bartels1995introduction,piegl1996nurbs}, Cubic Spline, B-spline, Bézier Curve, NURBS Curve, etc.
Then choose the minimum number of controlling nobs for that curve, keeping the rules of monotonicity. 

In Fig. \ref{fig:rl_warp}, we assume cubic splines with three knots; the central knot's position corresponds to the central pulse, and the edge knots' positions correspond to the edge pulses. The curves are symmetric around the central knot, and we will fix the warping function and roll-off factor values of the central knot. What remains is to control the edge knot positions and the amount of curvature. The curvature corresponds to the $T$ parameter in Eq. (\ref{eq:linprog}). The figure shows the spline segment between the central knot and one of the edge knots. Moreover, if we fix the spline curvature parameter, we will only have the position of edge knots as an RL state $s_1, s_2$, and incrementations as actions $a_1, a_2$. 

The oversampling ratio should also be an essential state that affects the computational complexity, power consumption, and out-of-band emission. Therefore, we consider it as a state $s_3$, and the action related to it as incrementing it positively or negatively $a_3$. In the case of the warping function, the RL spline is only to be used as a guide form which a rounding operation should determine the position of the knots, in Eq. (\ref{eq:splinewarp}), at $w^{-1}(n)$. The new knots should be at the nearest sample point grid, as shown in Fig. \ref{fig:rl_warp}. Finally, the number of pulses $N$ can be a parameter that contributes to optimizing the latency versus computational complexity compromise. 
We can summarize the proposed states and actions as:

 \begin{itemize}
  \item States: $s_1 : s_1 \in x_t,~ s_2 : s_2 \in [0, 1] ,~ s_3 : s_3 = V ,  ~ s_4 : s_4 = N$.
  \item Actions: $a_1= \pm T_s ,~a_2 = \pm \delta \alpha: \delta \alpha \ll 1 ,~ a_3=\pm 1,~ a_4=\pm 1$.
\end{itemize}
 
\section{Simulation \& Results}
% \begin{tabular}{@{}c@{}}Guard duration \\ (samples)\end{tabular}
% Guard duration (samples)

\begin{table}[h]
\begin{center}
\begingroup
\small
\begin{tabular}{||c c c c c c||} 
 \hline
    & \rotatebox{270}{IFFT size} & \rotatebox{270}{DFT size} & \rotatebox{270}{$[z_h~,~  z_t]$}& \rotatebox{270}{\begin{tabular}{@{}c@{}}CP length \\ (samples)\end{tabular}} & \rotatebox{270}{\begin{tabular}{@{}c@{}}Guard duration~~ \\ (samples)\end{tabular}} \\  
 \hline\hline
  ZT-DFT-s-OFDM$_{z=2}$ & 128 & 16 & [2,2] & - & 32 \\ 
 \hline
  ZT-DFT-s-OFDM$_{z=3}$ & 128 & 18 & [3,3] & - & 42\\
 \hline
   ZT-DFT-s-OFDM$_{z=4}$&128 & 20 & [4,4] & - & 51 \\ [0.2ex] 
 \hline
 CP-DFT-s-OFDM & 104 & 12 & - & 24 & 24\\ 
 \hline
 CP-OFDM & 104 & - & - & 24 & 24\\
 \hline
 Warped Symm. & - & - & [1,1] & - & 24\\
 \hline
 Warped Asymm. & - & - & [1,1] & - & 24\\
 \hline
\end{tabular}
\endgroup

\caption{Simulated Waveforms Parameters.}
\label{tab:params}
\end{center}
\end{table}

Based on the context of mMTC, we want to evaluate the proposed waveform in a harsh situation requiring well-contained symbols.
We compare different variations of ZT-DFT-s-OFDM, CP-OFDM, and CP-DFT-s-OFDM waveforms. DFT-s-OFDM-based schemes are chosen for comparison because of their similarity in the time-frequency domain occupancy with the proposed waveform; they have comparable PAPR values. Zero tail schemes are selected because the low tail power causes a reduced OOBE in the presence of time offsets, making it relative to the proposed and fitting the mMTC applications. %Note that we will not evaluate across any GFDM symbol because it has many variations regarding the pulse shapes and inter-pulse spacing. Moreover, any GFDM variation can have its corresponding warped version. Warped GFDM symbols will be left for a future study. 
Finally, we compare with CP-OFDM, which is always used as a baseline for any new waveform. The parameters of the simulated waveforms are listed in Table \ref{tab:params}. 

The roll-off factors profile of the warped symbols are taken from Table \ref{tab:problem}, and the inner $\alpha$ values for the asymmetric pulses are taken from Table \ref{tab:solution}. The warping parameters for the waveform of the symmetric pulses are taken from Table \ref{tab:solution}. In this evaluation, the asymmetric pulse waveform's warping parameters are the same as those used for the symmetric pulses. The difference between the two waveforms will be reflected in the lower OOBE for the asymmetric pulses.  

%The ZT-DFT-s-OFDM word consists  of 32 symbols (26 data symbols + 2 zeros-heads + 4 zero-tails). The warped word consists of 30 symbols (26 data symbols + 2 zeros-heads + 2 zero-tails), with fewer symbols due to the warping dilation effect. ZT-DFT-s-OFDM has an oversampling ratio of 4. Hence the time samples are 128, and the warped signal will have the same time duration $128 T_s$, as shown in Fig. \ref{fig:time}. 

%The pulses $\alpha$ factors follow the profiles shown in Fig. \ref{fig:alphaprofiles}; The $\alpha_1$ values for the pulses on the left edge are high, and they decrease as the pulses approach the inner part of the word. The same happens with $\alpha_2$ and the corresponding pulses on the right edge of the word. The warping function was determined to have the same spectral occupancy of the ZT-DFT-s-OFDM word.  
This section has spectral and temporal shape evaluations first. Then in the second subsection, bit error rate (BER) versus signal to interference noise ratio (SINR) evaluations, where interference is generated from adjacent symbols of the same type, in time and frequency. Finally, the third subsection evaluates the BER when the time domain interferer is offset in time and overlaps with the victim symbol. 

%First, we compare the spectral and time-domain containment. Then we evaluate the performance in the presence of delay-spread induced interference. Finally, the gain in the existence of a time offset is shown.
\subsection{Waveform Time-Frequency Containment}

\begin{figure}[ht]
\begin{subfigure}{0.48\textwidth}
         \centering
         \includegraphics[width=\textwidth]{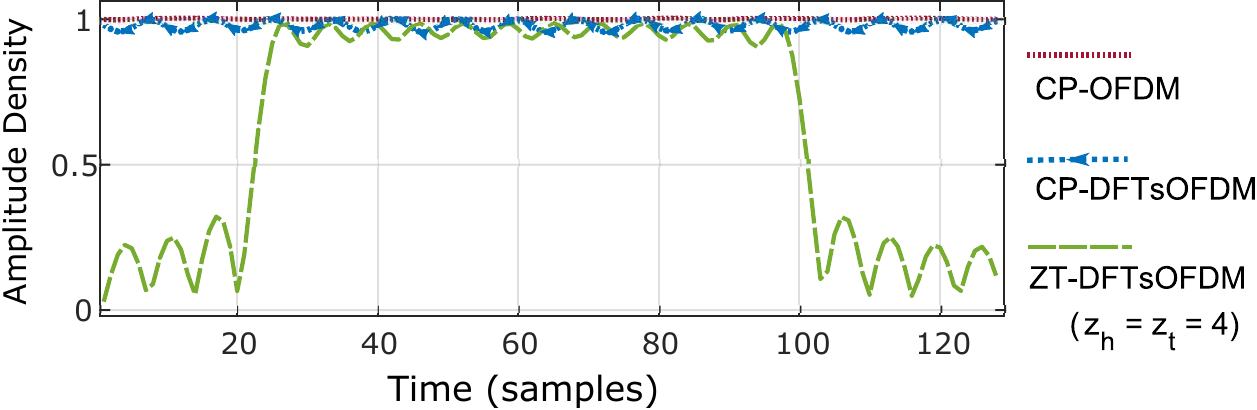}
         \caption{}
         \label{fig:time_1}
     \end{subfigure}
    \begin{subfigure}{0.48\textwidth}
         \centering
         \includegraphics[width=\textwidth]{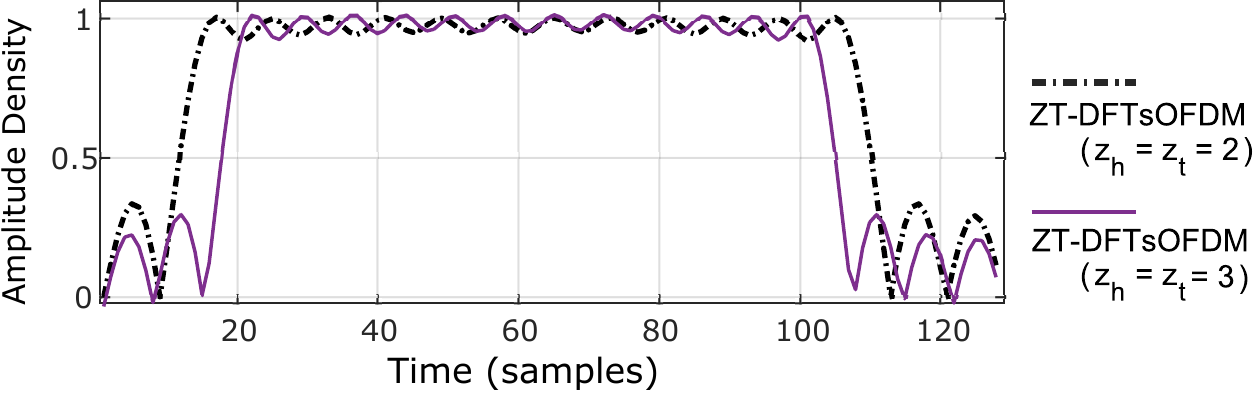}
         \caption{}
         \label{fig:time_2}
     \end{subfigure}
     \begin{subfigure}{0.48\textwidth}
         \centering
         \includegraphics[width=\textwidth]{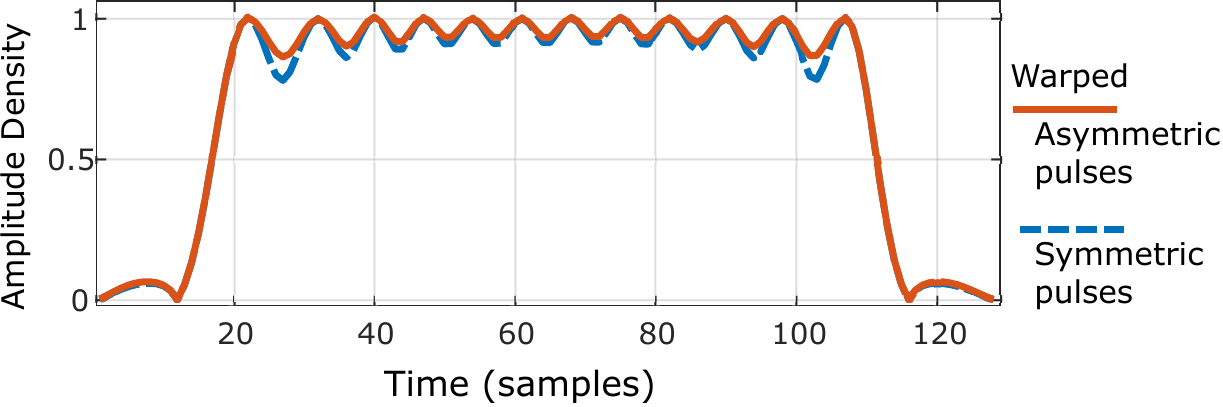}
         \caption{}
         \label{fig:time_3}
     \end{subfigure}
    %\includegraphics[width=.48\textwidth]{figs/time_density_.pdf}
    %\begin{center}
    \caption{Time domain amplitude density.}
    \label{fig:time}
\end{figure}

\begin{figure}[ht]
    \includegraphics[width=.45\textwidth]{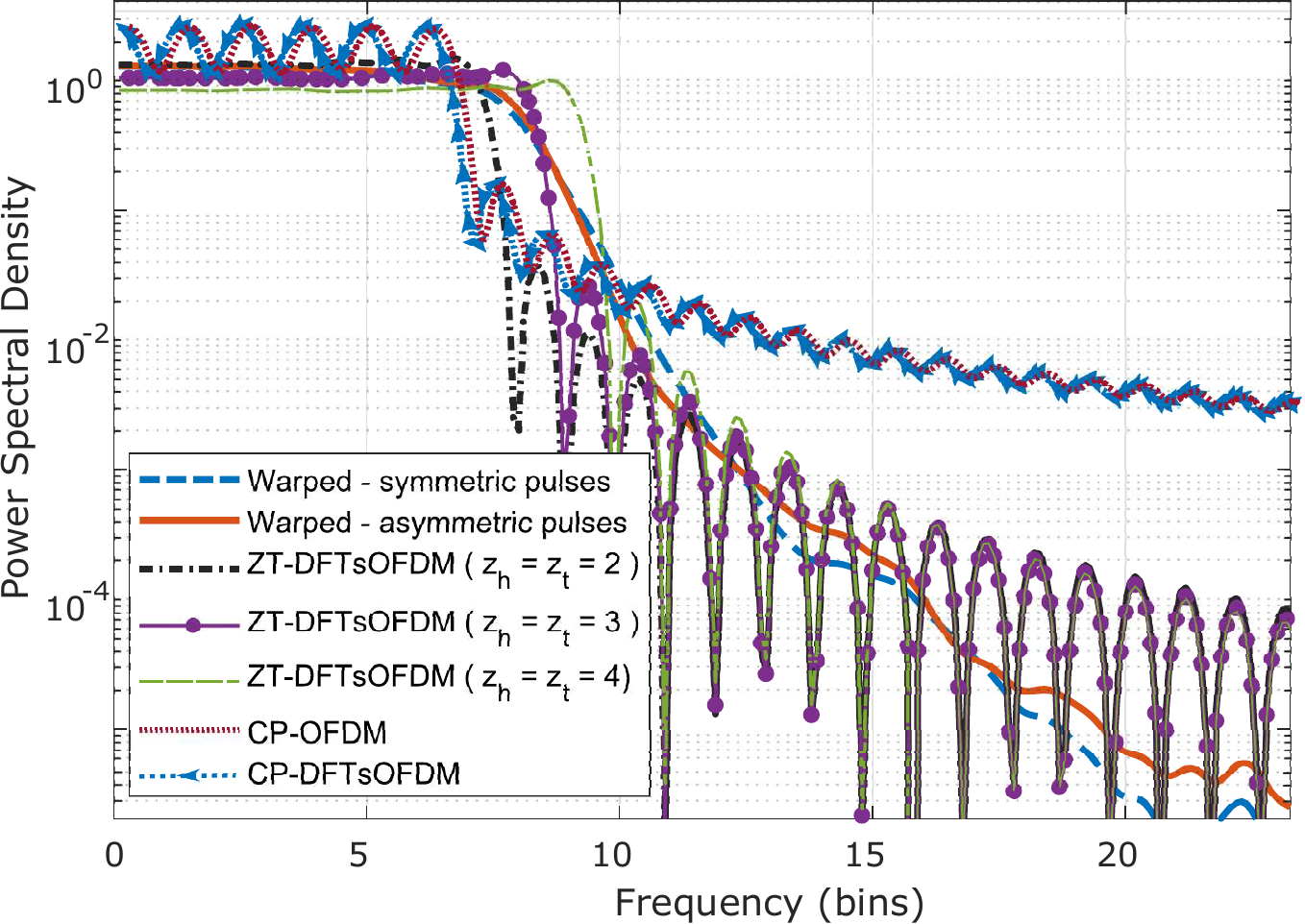}
    %\begin{center}
    \caption{Frequency domain power spectral density.}
    \label{fig:freq}
\end{figure}

\begin{figure}[ht]
    \includegraphics[width=.48\textwidth]{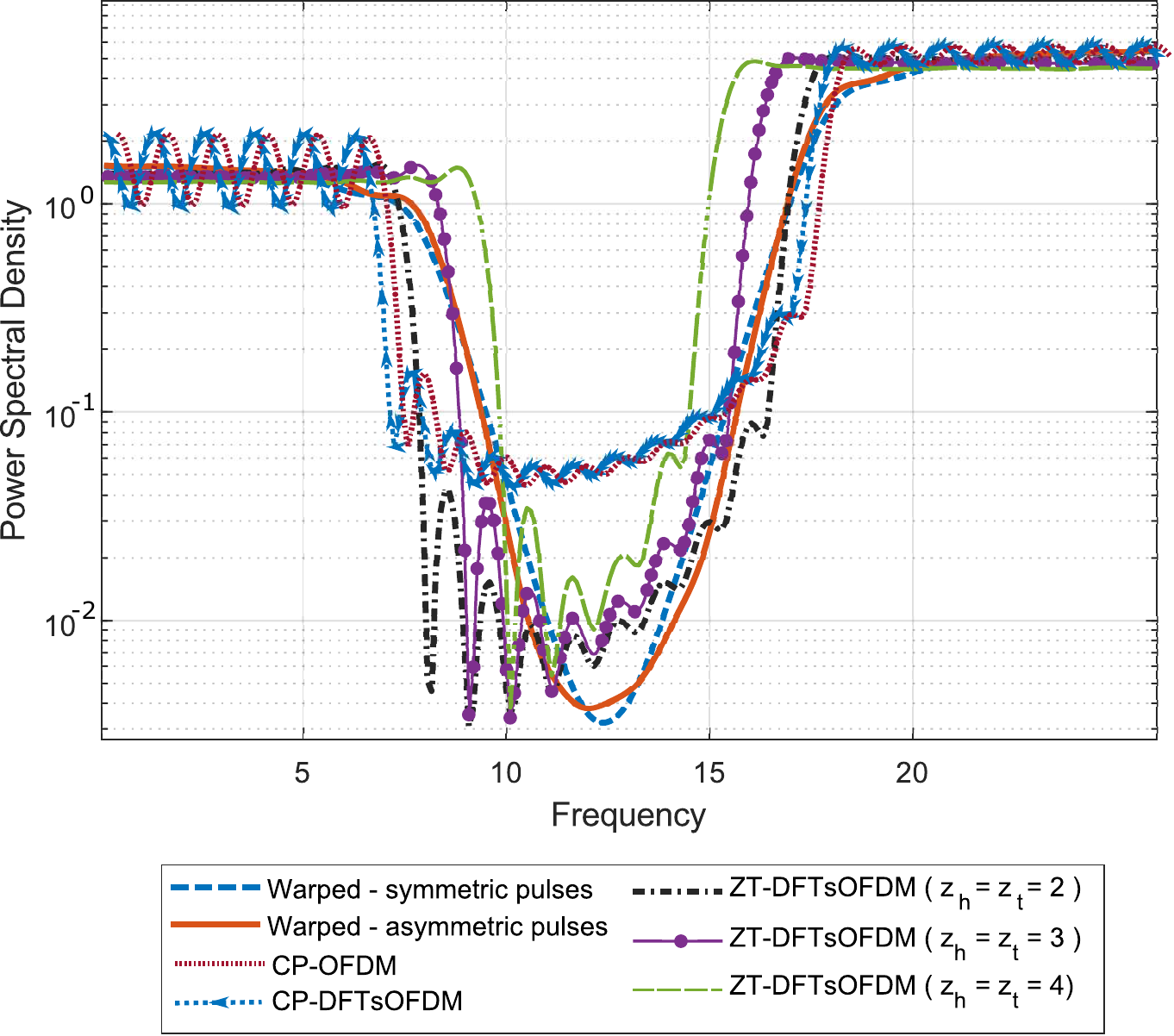}
    %\begin{center}
    \caption{Frequency domain power spectral density with interfering bands.}
    \label{fig:freq_intrfr}
\end{figure}

The symbols are compared in Fig. \ref{fig:time} for the time domain amplitude profile and Fig. \ref{fig:freq} for the spectral power profile. We can observe that the warped waveform in Fig. \ref{fig:time_3} is dilated in the time domain such that the 12 pulses occupy more time domain samples than the other waveforms, but it has lower power tails compared to ZT-DFT-s-OFDM. For ZT-DFT-s-OFDM in Fig. \ref{fig:time_1} and \ref{fig:time_2}, the tail power for the last zero decreases with increasing the number of guard zeros, but it does not reach the level of the warped waveforms. The symmetric and asymmetric pulses have almost the same amplitude density in the time domain.

\begin{figure}[ht]
\centering
    \includegraphics[width=.47\textwidth]{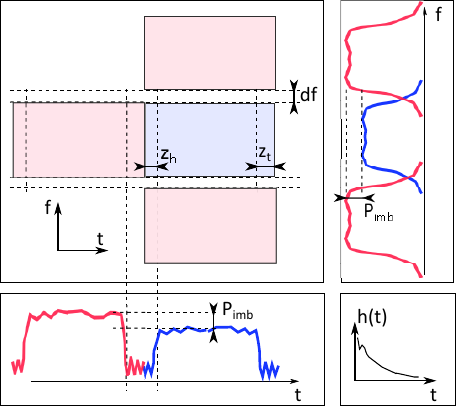}
    %\begin{center}
    \caption{Simulation Environment.}
    \label{fig:simsetup}
\end{figure}

\begin{figure}[ht]
    \includegraphics[width=.48\textwidth]{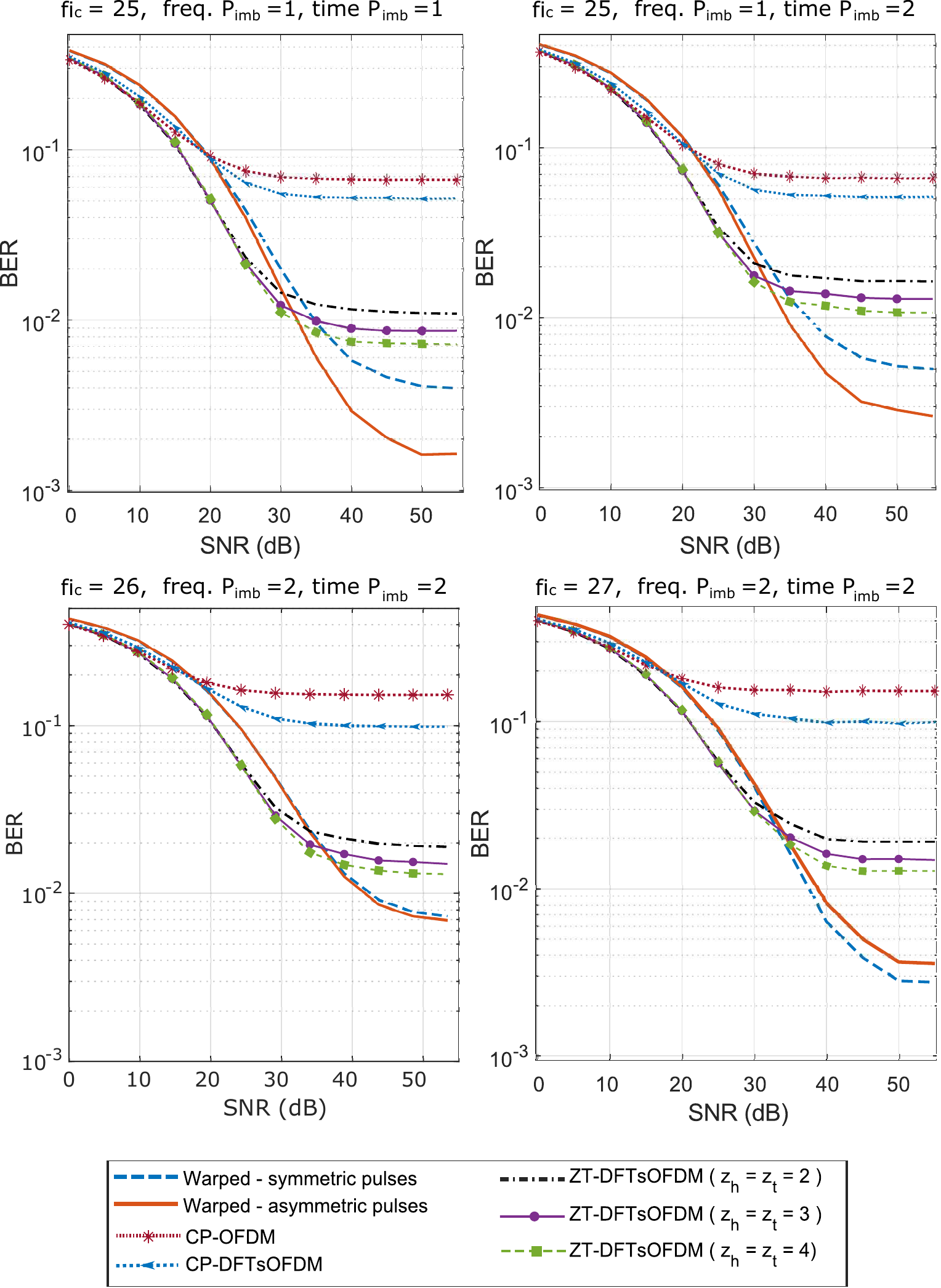}
    \caption{BER versus SNR for ZT-DFT-s-OFDM symbol and an equivalent warped symbol at different power imbalance and frequency spacing values.}
    \label{fig:bervssnr}
    %\end{center}
\end{figure}

The spectral power density of the CP-OFDM and CP-DFT-s-OFDM has the highest OOBE due to the rectangular window shape in the time domain. The zero tail-based waveforms have lower OOBE and their occupancy increases with increasing the gap duration in the time domain. This is because increasing the zero heads and tail $z_h$, $z_t$ for the same symbol duration squeezes more pulses in time. We can observe that the power dropoff for the warped waveforms can achieve lower than the conventional ZT-DFT-s-OFDM in case of choosing a low enough value of the optimization parameter $L$ in Eq. (\ref{eq:linprog})
For further clarification, we show in Fig. \ref{fig:freq_intrfr} the spectral shape of the interference from a higher power symbol to our victim symbol. The interferer symbols share the same center frequency.
We can observe that the warped waveforms have the lowest frequency domain intersymbol interference even with the shown 3 dB power imbalance.
 %Moreover, the gain in spectral containment due to using warped asymmetric pulse shapes versus the symmetric pulses with the same roll-off factor profile is shown in Fig. \ref{fig:freq}. 

\subsection{Delay Spread Effect on Performance}
In this subsection, we evaluate the BER performance based on the time-frequency resource grid mapping shown in Fig. \ref{fig:simsetup}. The evaluated waveform is affected by time and frequency domain interferers. Hence, the performance results are interference limited and will show a noise floor that corresponds to the interference-induced BER degradation at almost zero additive white Gaussian noise (AWGN). The channel model is a simple small-scale fading model with exponentially decaying power delay profile spread \cite{Bello}. $\tau_{rms}$ is the root means square (rms) delay coefficient. No large-scale fading is considered in this simulation. 

Time domain inter-symbol interference (ISI) between two ZT-DFT-s-OFDM symbols occurs when they are sent right after each other. ISI is present due to the concatenation of the two symbols. In the presence of a time-dispersive channel, the zero-tails from the first symbol leak power into the second symbol. Therefore, the second symbol's bit error rate (BER) decreases with the delay coefficient and the power imbalance $P_{imb}$ between the two symbols.

The interference in the frequency domain comes from similar symbols with center frequency shifted with the same amount for all the evaluated waveforms, which evaluates their spectral occupancies.

Zero-forcing frequency domain equalizer is used for the next BER versus signal-to-noise ratio (SNR) simulation. The High order quadrature amplitude modulation (512 QAM) modulation is used, and a root means square delay spread of $4 T_s$ is assumed. Also, the power ratio between the interferer and the evaluated symbol has two values of 0 dB and 3 dB. The results are summarized in Fig. \ref{fig:bervssnr}.

\begin{figure*}[t]
     \centering
     \begin{subfigure}[b]{0.37\textwidth}
         \centering
         \includegraphics[width=\textwidth]{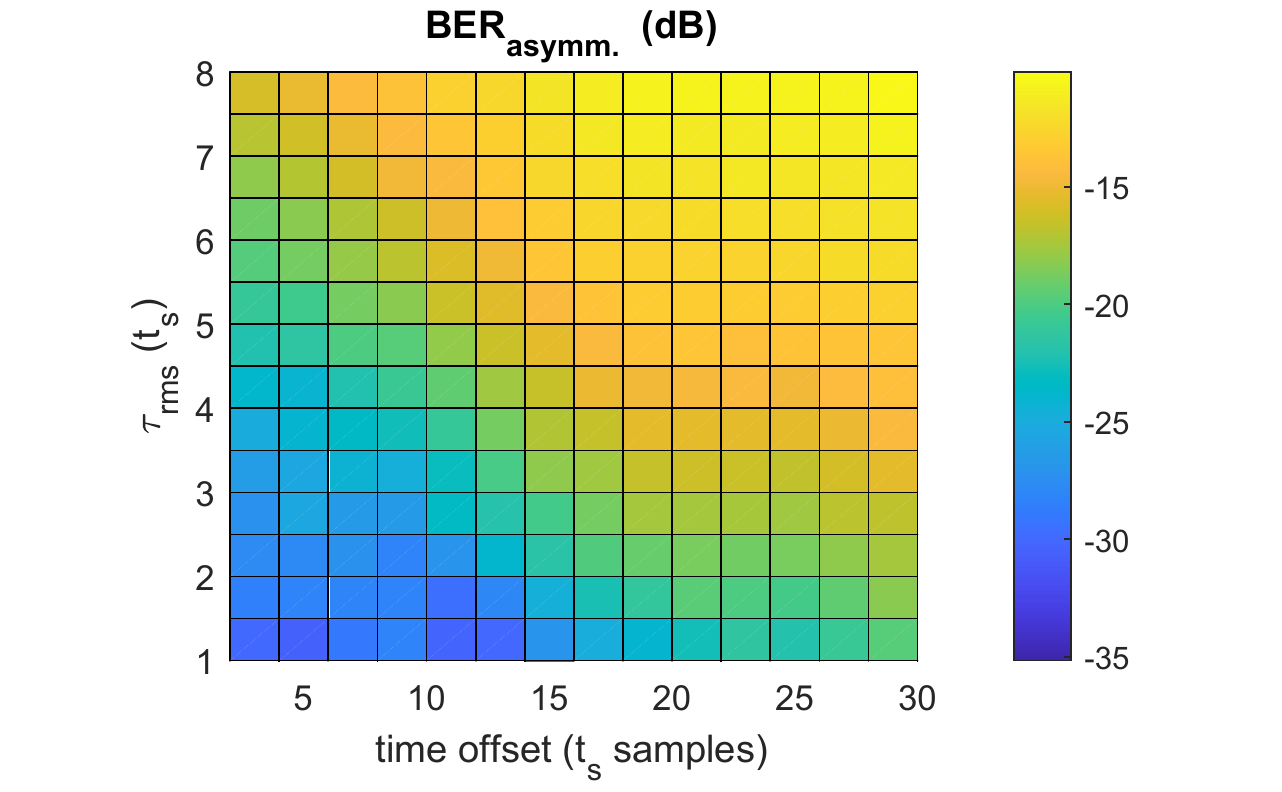}
         %\caption{$y=x$}
         \label{fig:Ca}
     \end{subfigure}
     \begin{subfigure}[b]{0.37\textwidth}
         \centering
         \includegraphics[width=\textwidth]{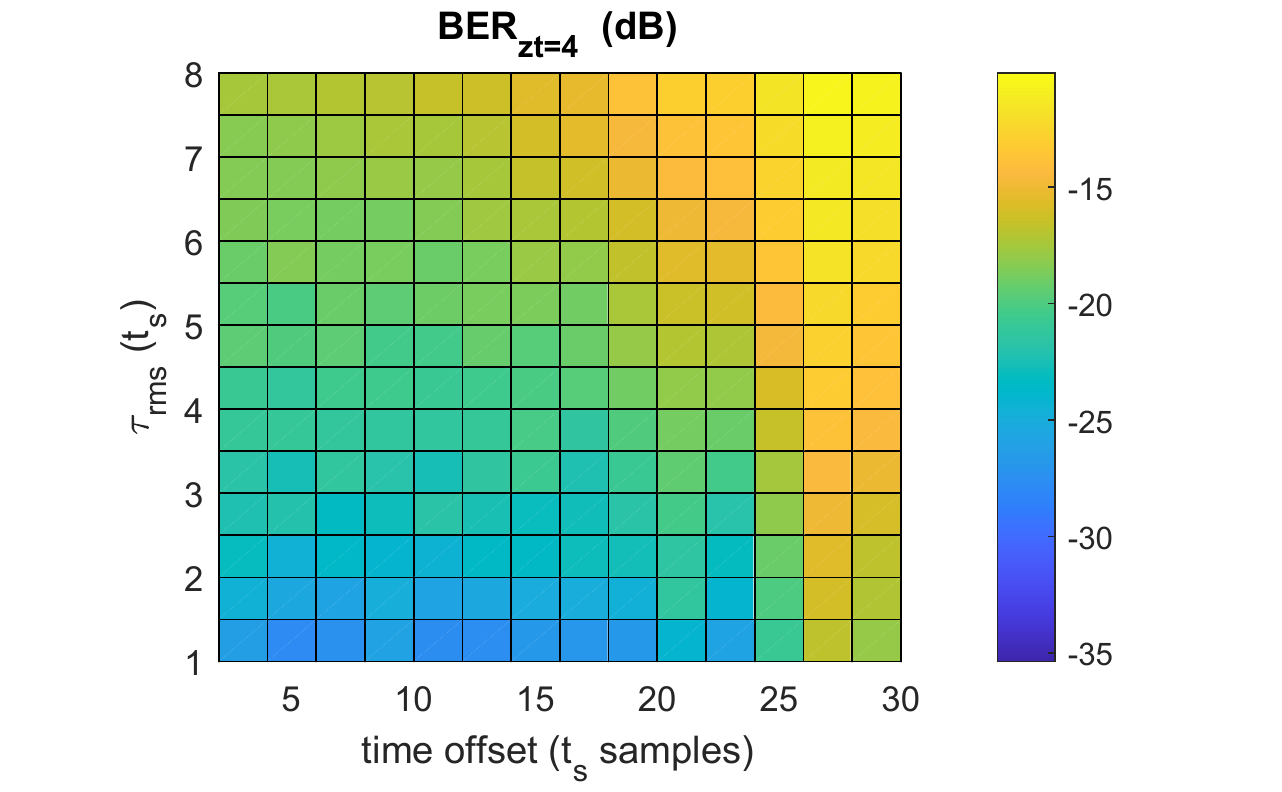}
         %\caption{$y=3sinx$}
         \label{fig:C4}
     \end{subfigure}
     \begin{subfigure}[b]{0.32\textwidth}
         \centering
         \includegraphics[width=\textwidth]{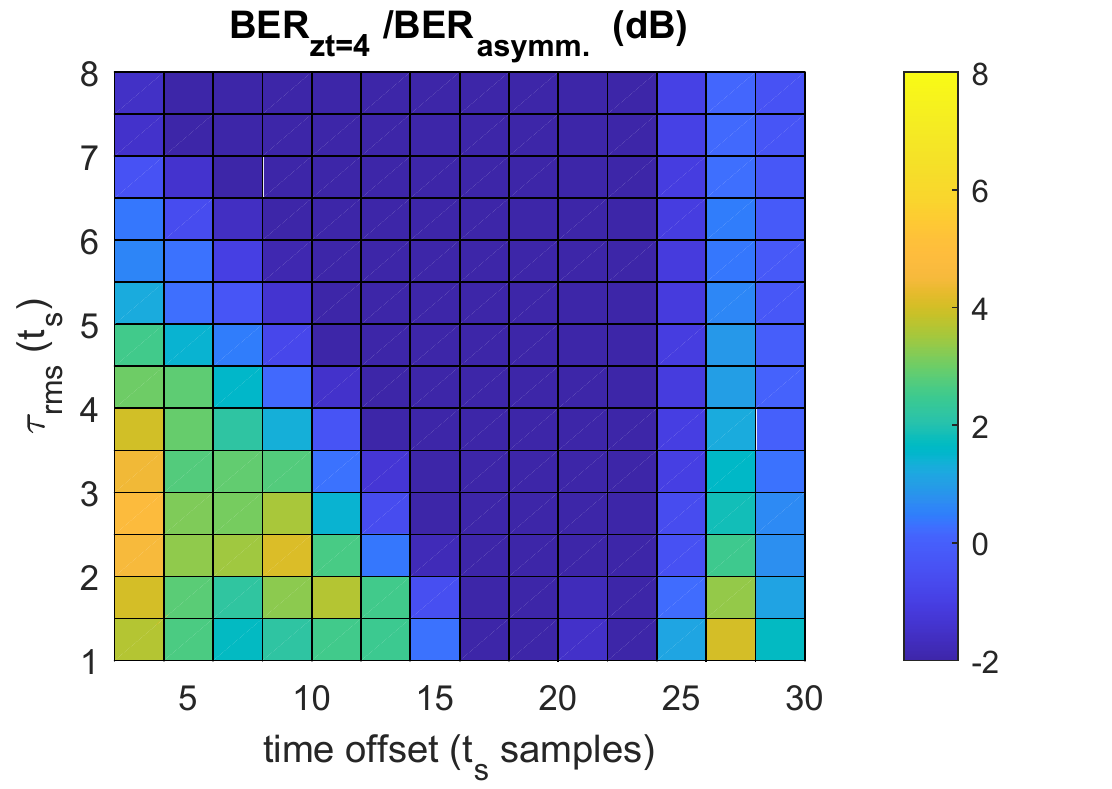}
         %\caption{$y=x$}
         \label{fig:C4a}
     \end{subfigure}
     \begin{subfigure}[b]{0.32\textwidth}
         \centering
         \includegraphics[width=\textwidth]{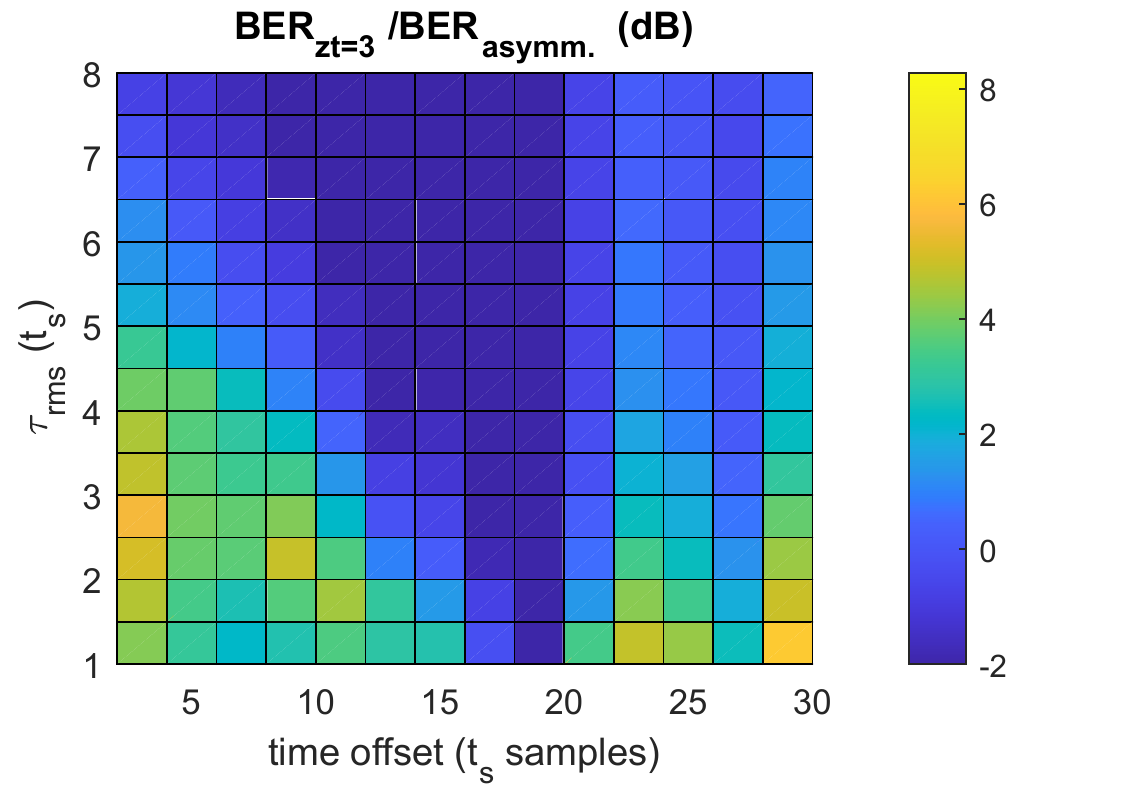}
         %\caption{$y=3sinx$}
         \label{fig:C3a}
     \end{subfigure}
     \begin{subfigure}[b]{0.32\textwidth}
         \centering
         \includegraphics[width=\textwidth]{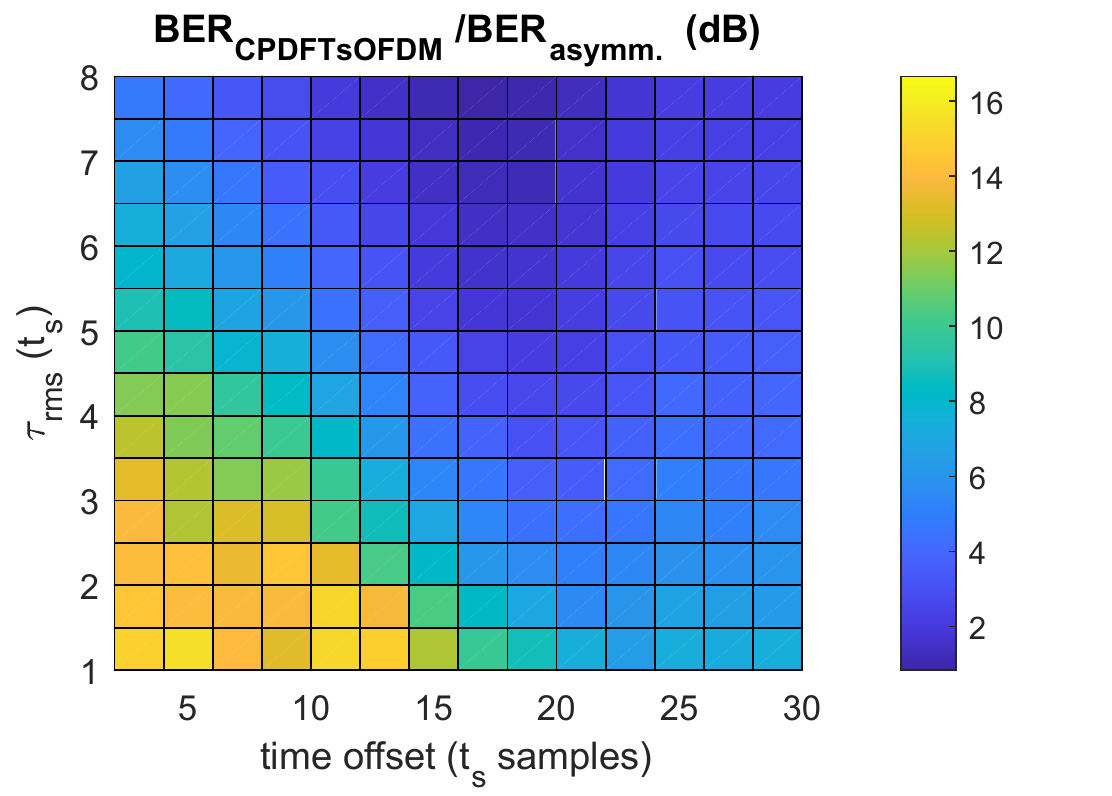}
         %\caption{$y=5/x$}
         \label{fig:Cca}
     \end{subfigure}
        \caption{BER floors, and performance gains in a loosely synchronized grid.}
        \label{fig:cc4a}
\end{figure*}
All the results shown next are specific to the generated warping function of the 12-pulse waveform and will be different for different warping functions or pulse numbers.
 We can observe from Fig. \ref{fig:bervssnr} 
 three different cases of power imbalances, and center frequency of interferer $fi_{c}$. The results are itemized as follows:
 \begin{itemize}
     \item For frequency $P_{imb}$ = 0 dB, time $P_{imb}$ = 0 dB, and $fi_{c}$= 25 bins: BER of warped asymmetric pulses has a performance floor  5 dB lower than the case of ZT-DFT-s-OFDM, zt= zh= 4. 
     \item For frequency $P_{imb}$ = 0 dB, time $P_{imb}$ = 3 dB, and $fi_{c}$=25 bins: BER of warped asymmetric pulses has a performance floor  7 dB lower than the case of ZT-DFT-s-OFDM, zt= zh= 4.
      \item For frequency $P_{imb}$ = 3 dB, time $P_{imb}$ = 3 dB, and $fi_{c}$= 26 bins: BER of warped asymmetric pulses has a performance floor  3.5 dB lower than the case of ZT-DFT-s-OFDM, zt= zh= 4. The asymmetric warped pulses have almost the same performance.
      \item For frequency $P_{imb}$ = 3 dB, time $P_{imb}$ = 3 dB, and $fi_{c}$= 27 bins: BER of warped symmetric pulses has a performance floor  7 dB lower than the case of ZT-DFT-s-OFDM, zt= zh= 4.
 \end{itemize}
  We notice that the symmetric warped pulses have a higher BER floor due to the higher spectral OOBE than the asymmetric pluses case. Except for the case of frequency $P_{imb}$= 3 dB, and $fi_{c}$= 26 bins, where the shape of the frequency domain drop causes the symmetric pulses to have better performance for the generated warping function. 
  Also, as we increase the zero gap duration for ZT-DFT-s-OFDM symbols, the tail's power decrease, and the BER is enhanced.    
 %that the BER of the warped waveform is lower than the BER of ZT-DFT-s-OFDM with around 10 dB for 16-QAM, 6 dB for 64-QAM, and 5 dB for 256-QAM.  

\subsection{Non-Perfect Synchronization Effect on Performance}

In this section, non-perfect synchronization is assumed to show the tolerance of the proposed method for a loosely synchronized resource grid. We determine the gain in BER for the warped waveform over the ZT-DFT-s-OFDM and CP-DFT-s-OFDM for 512-QAM modulation at SNR= 50 dB. A 0 dB power imbalance in time and frequency domains between the main symbol and the interferer symbols is assumed. The previously used symbols are reused for this subsection. 
The interferer symbol in the time domain is shifted and overlaps the evaluated symbol with a time offset ($t_s$ samples). In other words, the interferer's zero heads overlap the victim's zero tails. Also, there is a time dispersive channel with a root means square delay coefficient $\tau_{rms}$.

The BER floors are evaluated across the time offset and $\tau_{rms}$ for the warped asymmetric pulses symbol and ZT-DFT-s-OFDM $z_h=4$, and shown in the upper two plots of Fig. \ref{fig:cc4a}. Then we compare the noise floors of the warped asymmetric pulses symbol with ZT-DFT-s-OFDM $z_h=4, z_h=3$, and CP-DFT-s-OFDM. The lower three plots show the reduction in noise floor for the asymmetric pulses warped waveform.

The value of the gain is simulated over the range of positive time offset $[0,30 ]~ T_s$, and rms delay spread $[0,4]~ T_s$. The positive time offset causes ISI, degrading the BER performance. However, we observe a gain of up to 3 folds when compared with ZT-DFT-s-OFDM $z_h=4$, a gain of 5 folds when compared with  ZT-DFT-s-OFDM $z_h=3$, and more than 20 folds for the rectangular windowed shaped CP-DFT-s-OFDM. 
 These results were expected because time offset and delay spread increases the leakage of tails and decrease BER consequently. But this affects the warped waveform less than the conventional waveform because it is well-contained.  

\subsection{ {Design With Higher N}}
\begin{figure}[h]
\begin{center}
  \begin{subfigure}{0.23\textwidth}
         \centering
         \includegraphics[width=\textwidth]{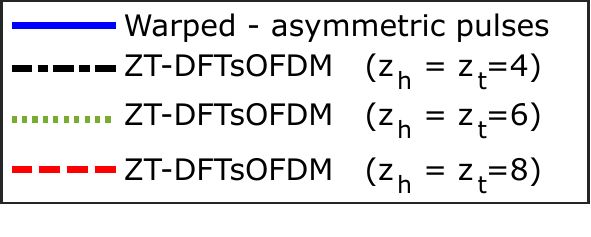}
         
         \label{fig:longlegend}
     \end{subfigure}
\begin{subfigure}{0.40\textwidth}
         \centering
         \includegraphics[width=\textwidth]{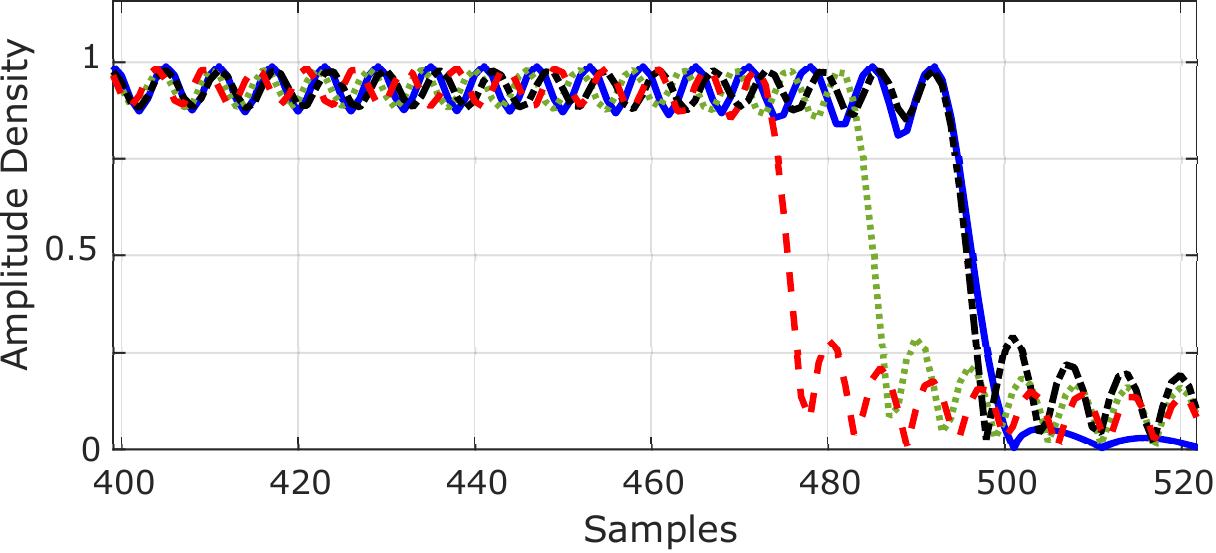}
        
         \label{fig:longtime}
     \end{subfigure}
    \begin{subfigure}{0.40\textwidth}
         \centering
         \includegraphics[width=\textwidth]{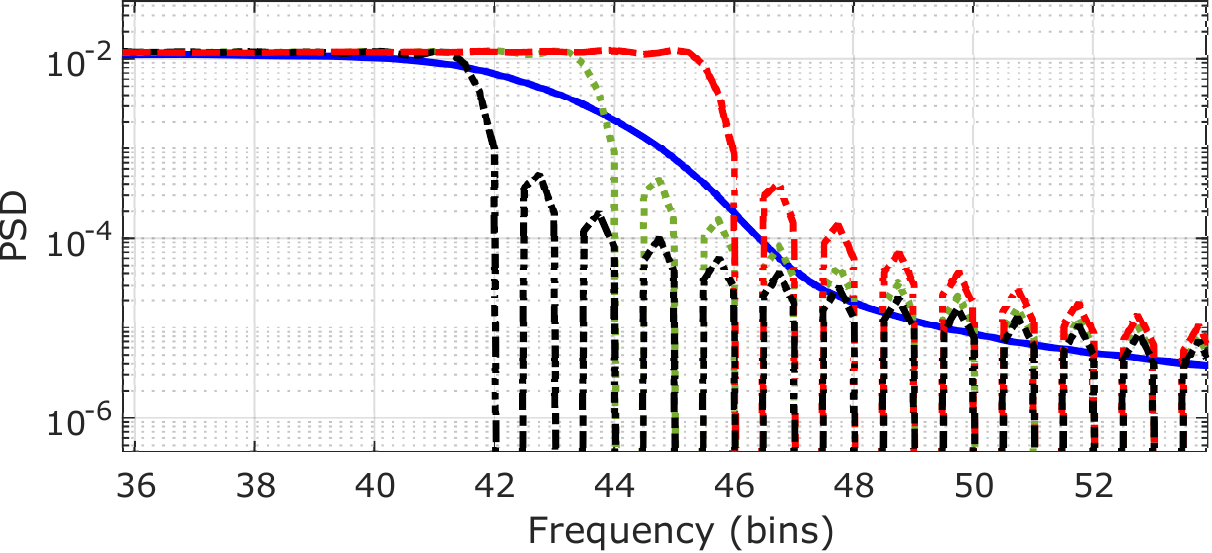}
         
         \label{fig:longfreq}
     \end{subfigure}

    \end{center}
    \caption{Time domain amplitude density and frequency domain PSD for N=76 Waveforms.}
    \label{fig:PSDlong}
\end{figure}

 {In this section, a warped waveform with a higher number of pulses is compared with various ZT-DFT-s-OFDM configurations. To simplify the demonstration, we only select the most competent waveforms based on our setup. We compare the proposed asymmetric pulse warped waveform with the ZT-DFT-s-OFDM variations using the same interference setup as above.
We provide a numeric example of the proposed mixed modulation architecture for the warped waveform, as described in Section V.C.1. The waveform being evaluated consists of 76 data symbols and corresponding pulses. The roll-off factors for this waveform are as follows:

$\alpha_{out}=$[1 , 0.48, 0.34, 0.27, 0.21, 0.17, 0.12, 0.09, 0.09, 0.09, ... , 0.09, 0.09, 0.09, 0.12, 0.17, 0.21, 0.27, 0.34, 0.48, 1 ],

$\alpha_{in}$=[0.22,  0.15, 0.09, 0.09, 0.09, ... , 0.09, 0.09, 0.15, 0.22].

The warping function is a spline piecewise function with anchor points as follows: }

\begin{equation*}
\begin{split}
w^{-1}  (n&;~n\in[1,~2,~…,~82  ])=  [1,~ 12,~ 23,~ 33, ~41, ~49, ~56,~ \\
 & 63, ~~69,~~ 75,~~ 81,~ 87, ~93, ~99,~   105,~111,~117,~ 123,~\\  
 &129,~ 135,~141,~ 147,~153, ~ 159, ~165,~ 171, ~ 177,~ 183, \\
 &189,~ 195,~201,~ 207,~213,~  219,~ 225,~ 231,~ 237,~   243,~\\ 
 & 249,~   255 ,~  261 ,~  267  ,~ 273   ,~279 ,~  285,~   291,~297,~   303,~ \\ 
 & 309,~   315,~   321,~   327,~   333,~   339,~ 345,~   351  ,~ 357  ,~ 363 ,\\ 
 & 369,~   375 ,~  381 ,~  387,~ 393   ,~399 ,~  405 ,~  411,~   417 ,~  423  ,\\ 
 &429 ,~  435,~ 441  ,~ 447 ,~  453,~   459 ,~  465 ,~  471 ,~  478,~485,~ \\ 
 & 492 ,~  501,~   511 ,~  522].   
\end{split}
\end{equation*}

   {The right-hand side of the diagram represents the positions of the samples to which the warped pulses are assigned. Noting that the first pulse is at $(n=4,w(n)=33)$ and the last pulse at $(n=79,w(n)=492)$. The warping function used here allows for E=20 for the filter bank scheme and V=6 for the parallel DFT-s-OFDM scheme. The sizes block of the DFT-s-OFDM scheme are: }
\begin{equation*}
\begin{split}
 & N_{DFT}=(N-2(E-z_e))=( 76-2(20-14))=64 \\
 & N_{window}=2N_{DFT}=128  \\
 & N_{IFFT}=(N-2(E-z_e))\times V= 384   ,
\end{split}
\end{equation*}
 {where $z_e$ denotes the zero inputs in the DFT-s-OFDM branch used to account for overlap between the two schemes and to create space for the tails of the inner pulses. The output spectrum of the DFT is extended by replication to accommodate for the windowing. %Another way to do this is to 
Alternatively, we can use a DFT of size $2(N-2(E-z_e))$ but with inputs spaced by a zero between each symbol. We note that the size of the IFFT vector is less than the filter bank output of size 522. The IFFT output will be zero-padded and positioned where the pulses should fit between the edge pulses.

The length of the warped waveform in samples is 522. Therefore, we compare it with ZT-DFT-s-OFDM waveforms of the same length. The DFT size is $(N+z_h+z_t )$, and the IFFT size is 522. 
} 

We illustrate the zero tails and the out-of-band frequency domain occupation for the simulated variations in Fig. \ref{fig:PSDlong}. We observe that as we increase the zero tails of the ZT-DFT-s-OFDM symbols, it takes up more bandwidth. This is because packing more pulses in the same duration leads to a compromise between the time and frequency occupancy.
\begin{figure}[ht]
\begin{subfigure}{0.24\textwidth}
         \centering
         \includegraphics[width=\textwidth]{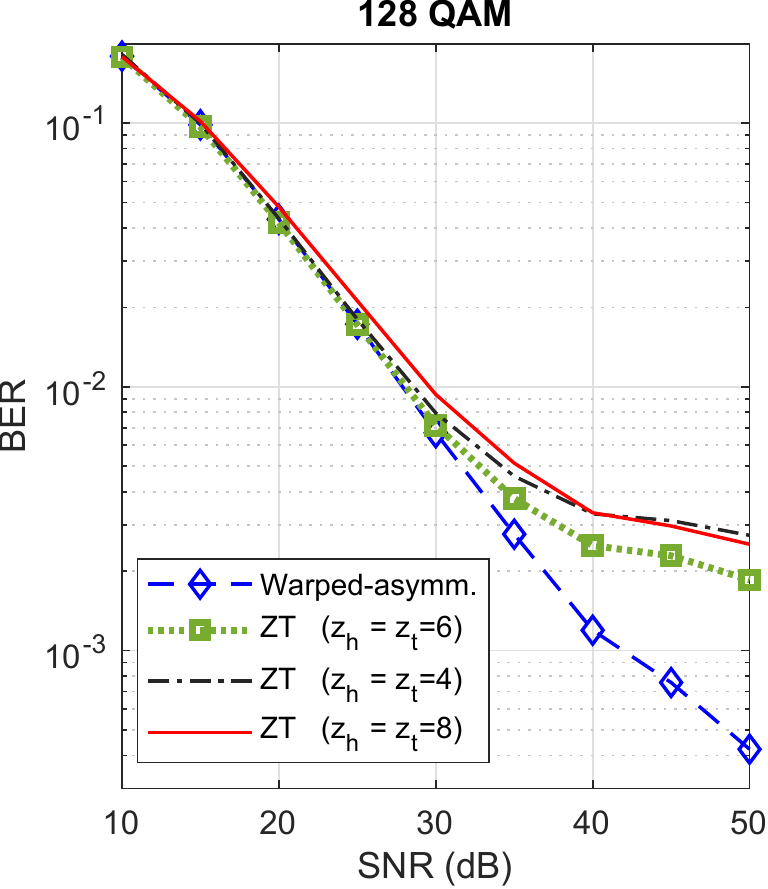}
         \caption{}
         \label{fig:long128}
     \end{subfigure}
    \begin{subfigure}{0.24\textwidth}
         \centering
         \includegraphics[width=\textwidth]{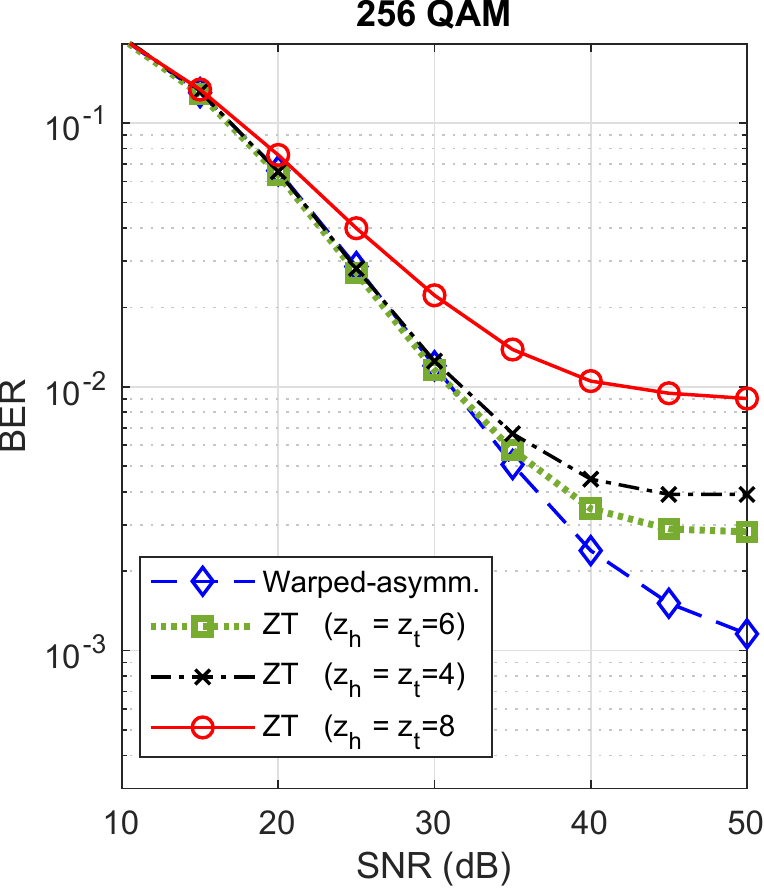}
         \caption{}
         \label{fig:long256}
     \end{subfigure}
   
    %\includegraphics[width=.48\textwidth]{figs/time_density_.pdf}
    %\begin{center}
    \caption{BER versus SNR for (N=76) warped asymmetric pulse symbol and ZT-DFT-s-OFDM (ZT), with different zero tails at different modulation orders.}
    \label{fig:BERlong}
\end{figure}

In Fig. \ref{fig:BERlong}, we evaluate the BER performance based on the time-frequency resource grid mapping shown in Fig. \ref{fig:simsetup}, which is the same as in Section VI.B. We use a delay spread of $\tau_{rms}=6~T_s$, time and frequency power imbalances of 0 dB each, and $fi_c=$93 bins. We compare the proposed warped asymmetric pulses with ZT-DFT-s-OFDM. We observe that the proposed waveform has lower BER floors because it is less affected by side interferers in time and frequency domains. This is valid for the same time-frequency occupancy assigned to both of the competing waveforms.

\section{Conclusion}
In this paper, a novel time-frequency axis warped waveform is proposed for low-power, well-contained, and synchronization-relaxed mMTC, IoT, and sensor network applications, where symbols are in short bursts with high QAM orders. Our waveform has low PAPR because it is based on the SC-OFDM scheme; it is well-contained in time and frequency domains, thanks to the proposed pulse shapes and axis warping procedure. 
 We use asymmetric RC pulse shapes to increase spectral containment, where a low power tail (high $\alpha$) is used toward the edges and a high power tail (low $\alpha$) is used toward the middle.
%Asymmetric RC pulse shapes are proposed to increase spectral containment. An asymmetric RC pulse shape is used with low power tails (high $\alpha$) toward the edges and high power tails (low $\alpha$) toward the middle.
A warping operation is used to dilate the symbol in the time domain so that all the pulses can occupy the same spectral window. 

%Finally, our waveform is evaluated and compared to ZT-DFT-s-OFDM in the delay spread ISI scenario and time offset ISI scenario. In both cases, gains by our waveform are observed for different power imbalances, frequency domain spacing, and high QAM orders due to the well-containment of the warped waveforms.  
We evaluate our waveform and compare it to ZT-DFT-s-OFDM in the delay spread ISI scenario and time offset ISI scenario. In both cases, we observe gains for different power imbalances, frequency domain spacing, and high QAM orders due to the well-containment of the warped waveforms.

As a final thought, the results in this paper are specific to the warping functions and roll-off factors profiles generated for the 12-pulse, and 76-pulse symbols. We chose a 12-pulse symbol because it is more presentable on paper with the length of roll-off and warping functions. Then we design and evaluate a longer warped waveform to provide a numeric example for the lower complexity mixed modulation scheme. %giving a numeric example for the lower complexity mixed modulation scheme.  %Also, for future studies, time-frequency warping for GFDM symbols with different pulse shapes and inter-pulse spacings can be explored. This will add more degrees of freedom to the waveform-shaping design space. 

\section*{Acknowledgment}
This material is based upon work supported in part by the U.S. Department of Energy, Office of Science, Office of Advanced Scientific Computing Research under Award Number DE-SC0023023. This work was supported in part by the U.S. National Science Foundation under Grants 1923295.
\ifCLASSOPTIONcaptionsoff
  \newpage
\fi
\bibliographystyle{IEEEtran}
\bibliography{Bibliography}
\vskip 0pt plus -1fil
\begin{IEEEbiography}[{\includegraphics[width=1in,height=1.25in,clip,keepaspectratio]{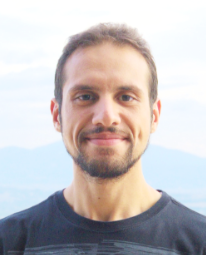}}]{Mostafa Ibrahim } received his B.Sc. degree in Electronics \& Electrical Communication Engineering from Ain-Shams University, Egypt, in 2010, M.Sc. degree in Electrical Engineering from Istanbul Medipol University, Turkey, in 2017. From 2018 to 2020, he led the implementation and development of a Cell Broadcast Center (CBC) (Project Initiator: BTK Turkish Information and Communication Technologies Authority). He took part in the Cell Broadcast Service deployment in two of Turkey's mobile network operators. Currently, he is working as a graduate research assistant at the School of Electrical and Computer Engineering, Oklahoma State University, Stillwater, OK, USA. His research interests include distributed management in wireless communication systems, beyond 5G waveform design, and air-ground channel modeling and measurements. \end{IEEEbiography}
\vskip 0pt plus -1fil
\begin{IEEEbiography}[{\includegraphics[width=1in,height=1.25in,clip,keepaspectratio]{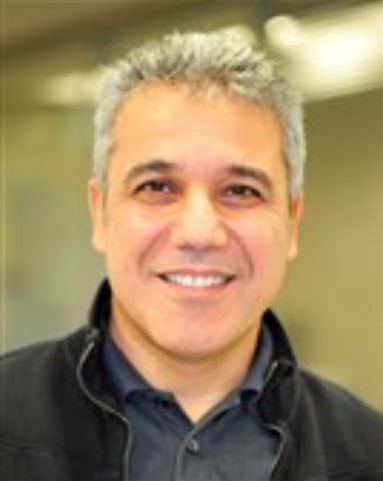}}]{Hüseyin Arslan } %(S’95–M’98–SM’04–F’15) received the B.S. degree in electrical and electronics engineering from the Middle East Technical University, Ankara, Turkey, in 1992, and the M.S. and Ph.D. degrees in electrical engineering from the Southern Methodist University, Dallas, TX, USA, in 1994 and 1998, respectively. From January 1998 to August 2002, he was with the research group of Ericsson Inc., Charlotte, NC, USA, where he was involved with several projects related to 2G and 3G wireless communication systems. He has worked as a Part Time Consultant for various companies and institutions including Anritsu Company (Morgan Hill, CA, USA), The Scientific and Technological Research Council of Turkey. He is currently a Professor of Electrical Engineering with the University of South Florida, Tampa, FL, USA, and the Dean with the College of Engineering and Natural Sciences, Istanbul Medipol University, ˙Istanbul, Turkey. 
%His research interests are related to advanced signal processing techniques at the physical and medium access layers, with cross-layer design for networking
%adaptivity and Quality of Service (QoS) control. He has served as Technical Program Committee Chair, Technical Program Committee Member, Session and
%Symposium Organizer and Workshop Chair in several IEEE conferences.
(IEEE Fellow, IEEE Distinguished Lecturer, Member of Turkish Academy of Science) received his BS degree from the Middle East Technical University (METU), Ankara, Turkey in 1992; his MS and Ph.D. degrees were received respectively in 1994 and 1998 from Southern Methodist University (SMU), Dallas, TX. He was with the research group of Ericsson between January 1998 and August 2002, and then at the University of South Florida, where he was a Professor. Since 2013, he has been working as the Dean of the School of Engineering and Natural Sciences at Istanbul Medipol University. He has worked as a part-time consultant for various companies and institutions including Anritsu Company, Savronik Inc., and The Scientific and Technological Research Council of Turkey. He was also the founding Chairman of The Board Of Directors of ULAK Communication company, which is the Turkish telecom equipment provider. Dr. Arslan’s research interests are related to advanced signal processing techniques, wireless PAN/LAN/MANs, fixed wireless access, aeronautical networks, in vivo networks, and wireless sensors networks. His current research interests are on 5G and beyond radio access technologies, physical layer security, interference management, cognitive radio, small cells, and underwater acoustic communications. Dr. Arslan has made scholarly contributions in various arenas and his research has created 55 issued patents and more than 75 pending patents, 51 book chapters, 3 edited books and one text book, more than 180 peer-reviewed journal papers (mostly IEEE journals) and more than 260 peer-reviewed IEEE conference papers. He has graduated 26 PhD students and 24 MS students.
\end{IEEEbiography}
% {{\includegraphics[width=25mm,height=32mm,clip,keepaspectratio]{a.eps}}}
\begin{IEEEbiography}[{\includegraphics[width=1in,height=1.25in,clip,keepaspectratio]{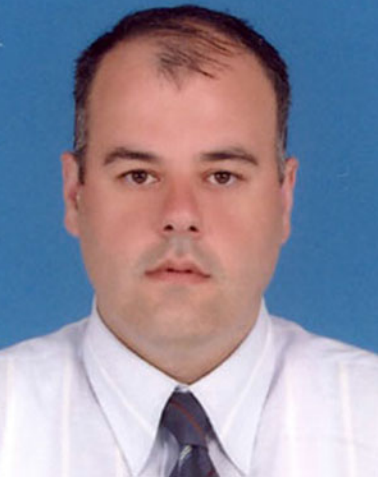}}]{Hakan Ali Cirpan } (Member, IEEE) received the B.Sc. degree in electrical engineering from Uludag University, Bursa, Turkey, in 1989, the M.Sc. degree in electrical engineering from Istanbul University, Istanbul, Turkey, in 1992, and the Ph.D. degree in electrical engineering from the Stevens Institute of Technology, Hoboken, NJ, USA, in 1997. From 1995 to 1997, he was a Research Assistant with the Stevens Institute of Technology, working on signal processing
algorithms for wireless communication systems. In 1997, he joined the Department of Electrical and Electronics Engineering, Istanbul University. In 2010, he joined the Department of Electronics and Communication Engineering, Istanbul Technical University. His current research interests include machine learning, signal processing, and communication concepts with specific attention to channel estimation and equalization algorithms for future wireless systems.\end{IEEEbiography}
\vskip 0pt plus -1fil
\begin{IEEEbiography}[{\includegraphics[width=1in,height=1.25in,clip,keepaspectratio]{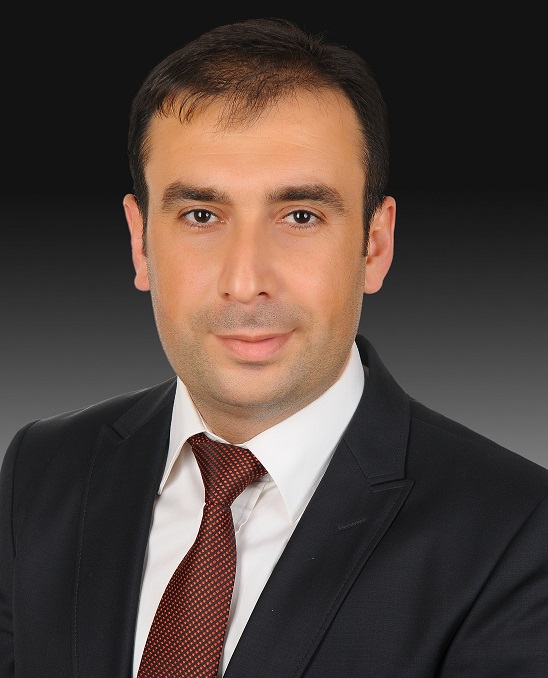}}]{Sabit Ekin (SM’21)} received his Ph.D. degree in Electrical and Computer Engineering from Texas A\&M University, College Station, TX, USA, in 2012. He has four years of industrial experience as a Senior Modem Systems Engineer at Qualcomm Inc., where he received numerous Qualstar awards for his achievements and contributions to cellular modem receiver design. He is currently an Associate Professor of Engineering Technology and Electrical \& Computer Engineering at Texas A\&M University. Prior to this, he was an Associate Professor of Electrical and Computer Engineering at Oklahoma State University. His research interests include the design and analysis of wireless systems, encompassing mmWave and terahertz communications from both theoretical and practical perspectives, visible light sensing, communications and applications, noncontact health monitoring, and Internet of Things applications.
\end{IEEEbiography}

\end{document}